\documentclass[journal]{IEEEtran}

\usepackage{amsfonts}
\usepackage{amssymb}
\usepackage{amsthm}
\usepackage{amsmath,amsfonts,amssymb}
\usepackage[dvips]{graphicx}
\usepackage{subfigure}
\usepackage{verbatim}
\usepackage{setspace}
\usepackage{bm}
\usepackage{algorithmic} 
\usepackage[ruled,vlined]{algorithm2e}
\usepackage{cite}

\usepackage{changepage}
\usepackage{pdfpages}
\usepackage{color}
\newcommand*{\dif}{\mathop{}\!\mathrm{d}}
\allowdisplaybreaks

\newcommand{\figwidth}{7.8}
\IEEEoverridecommandlockouts

\begin{document}
\title{Dynamic Beam Coverage for Satellite Communications Aided by Movable-Antenna Array}
\author{Lipeng Zhu, ~\IEEEmembership{Member,~IEEE,}
	    Xiangyu Pi,~\IEEEmembership{Graduate Student Member,~IEEE,}
		Wenyan Ma,~\IEEEmembership{Graduate Student Member,~IEEE,}
		Zhenyu Xiao,~\IEEEmembership{Senior Member,~IEEE,}
		and Rui Zhang,~\IEEEmembership{Fellow,~IEEE}
	\vspace{-0.5 cm}
	\thanks{L. Zhu and W. Ma are with the Department of Electrical and Computer Engineering, National University of Singapore, Singapore 117583 (e-mail: zhulp@nus.edu.sg, wenyan@u.nus.edu).}
	\thanks{X. Pi and Z. Xiao are with the School of Electronic and Information Engineering, Beihang University, Beijing, China 100191. (e-mail: pixiangyu@buaa.edu.cn, xiaozy@buaa.edu.cn)}
	\thanks{R. Zhang is with School of Science and Engineering, Shenzhen Research Institute of Big Data, The Chinese University of Hong Kong, Shenzhen, Guangdong 518172, China (e-mail: rzhang@cuhk.edu.cn). He is also with the Department of Electrical and Computer Engineering, National University of Singapore, Singapore 117583 (e-mail: elezhang@nus.edu.sg).}
}

\maketitle


\begin{abstract}
	The low-earth orbit (LEO) satellite network has been recognized as a promising technology to enable the ubiquitous coverage and massive connectivity for future sixth-generation (6G) mobile communications. Due to the ultra-dense constellation, efficient beam coverage and interference mitigation are crucial to LEO satellite communication systems, while the conventional directional antennas and fixed-position antenna (FPA) arrays both have limited degrees of freedom (DoFs) in beamforming to adapt to the time-varying coverage requirement of terrestrial users. To address this challenge, we propose in this paper utilizing movable antenna (MA) arrays to enhance the satellite beam coverage and interference mitigation. Specifically, given the satellite orbit and the coverage requirement within a specific time interval, the antenna position vector (APV) and antenna weight vector (AWV) of the satellite-mounted MA array are jointly optimized over time to minimize the average signal leakage power to the interference area of the satellite, subject to the constraints of the minimum beamforming gain over the coverage area, the continuous movement of MAs, and the constant modulus of AWV. The corresponding continuous-time decision process for the APV and AWV is first transformed into a more tractable discrete-time optimization problem. Then, an alternating optimization (AO)-based algorithm is developed by iteratively optimizing the APV and AWV, where the successive convex approximation (SCA) technique is utilized to obtain locally optimal solutions during the iterations. Moreover, to further reduce the antenna movement overhead, a low-complexity MA scheme is proposed by using an optimized common APV over all time slots. Simulation results validate that the proposed MA array-aided beam coverage schemes can significantly decrease the interference leakage of the satellite compared to conventional FPA-based schemes, while the low-complexity MA scheme can achieve a performance comparable to the continuous-movement scheme. 
\end{abstract}
\begin{IEEEkeywords}
	Satellite communications, low earth orbit (LEO) satellite, movable antenna (MA), beamforming, interference mitigation.
\end{IEEEkeywords}

%
\IEEEpeerreviewmaketitle

\section{Introduction}
\IEEEPARstart{T}{he} sixth-generation (6G) mobile communication systems are expected to support massive communication and ubiquitous connectivity \cite{dang2020should,wang2023road,ITU6G}. It is forecast that the number of connected wireless devices will exceed 50 billions worldwide by 2030 \cite{s21165284}. However, conventional terrestrial network infrastructures face serious challenges in fulfilling the coverage requirement for the ever-increasing number of wireless terminals \cite{Ghena2019challenge,Abouzaid2021sky}. On one hand, since the coverage range of a terrestrial base station (BS) is limited, it is cost prohibitive to deploy dense BSs in remote and/or sparsely populated areas, such as deserts, forests, highlands, and oceans. On the other hand, the terrestrial infrastructures are fragile to natural disasters, which render digital divide in the regions with broken BSs. To overcome these challenges, satellite communications have recently experienced a significant resurgence of interest \cite{Di2019LEO,Liu2021LEO,xiao2022LEO}. Due to the technological advancements and cost reductions, the deployment of ultra-dense low earth orbit (LEO) satellite constellations has been recognized as a promising solution to realize the global coverage and meet the high-throughput communication requirement of terrestrial devices. During the past few years, a number of practitioners have started their projects to establish LEO access networks, such as Starlink, OneWeb, and Lightspeed, which pave the way to fostering an Internet of everything (IoE) \cite{Akan2023IoE}.

Due to the long propagation distance and high operating frequency bands of satellite-ground links, LEO satellites are usually equipped with directional antennas or antenna arrays with high beam gains to compensate for the significant path loss \cite{Yahya2015antenna,Rao2015antenna}. A directional antenna usually has a fixed radiation pattern, which can only adjust its beam pointing via mechanical rotation. In contrast, an antenna array can synthesize different beam patterns by tuning the signals' phases and/or amplitudes over all antenna elements \cite{He2021review,Moon2019phased,Ning2023widebeam}. Thanks to such beamforming capabilities, antenna arrays can outperform directional antennas in terms of flexible beam coverage for satellite communications and have been widely adopted in existing systems \cite{Moon2019phased,Ran2023dual}. In general, the satellite-mounted antenna arrays can be mainly classified into two categories, i.e., uniform array and sparse/non-uniform array \cite{He2021review,Moon2019phased,Ning2023THzbeam}. A uniform array has regular geometry and thus it is easy to implement beamforming and beam scanning. However, the dense deployment of antennas may lead to severe coupling effect and the integrated circuits encounter great challenges in heat dissipation, especially in the space environment without cross-ventilation. Thus, the sparse/non-uniform array was proposed by enlarging the inter-antenna spacing to reduce the coupling effect \cite{Vigan2010sparse,Bucci2014sparse}. However, sparse arrays usually suffer from obstinate sidelobes when performing beamforming, which may cause severe interference leakage to terrestrial users.  

Since the number of satellites in an LEO constellation is extremely large, the interference leakage of satellites may dramatically deteriorate the network performance \cite{xiao2022LEO,Deng2020much,Deng2021many}. Thus, efficient interference mitigation between the beams of different satellites is essential for LEO communication systems. However, conventional uniform and/or sparse antenna arrays both employ fixed-position antennas (FPAs) \cite{He2021review,Moon2019phased,Vigan2010sparse,Bucci2014sparse}, where the fixed array geometry uniquely determines the steering vectors over different wave directions. As the satellite travels in an orbit, the directions of the coverage and interference areas relative to the antenna array keep changing over time. Since conventional FPA arrays cannot change the inherent correlation between the steering vectors over coverage and interference directions, only the antenna weights can be reconfigured to adapt to the varying coverage requirement of terrestrial users, which thus limits the degree of freedom (DoF) in beamforming for interference mitigation in LEO satellite networks.

To overcome the limitation of conventional FPA arrays, we propose in this paper utilizing movable antenna (MA) arrays to enhance the dynamic beam coverage for LEO satellite communications. Specifically, MA refers to the antenna with the capability of local movement at the transmitter/receiver \cite{zhu2023MAMag}, which is also known as fluid antenna system (FAS) \cite{zhu2024historical,wong2022bruce}. By integrating multiple MAs to form an array, the array geometry can be efficiently reconfigured via antenna movement. By jointly optimizing the antenna position vector (APV) and antenna weight vector (AWV), more flexible beamforming can be achieved by MA arrays as compared to traditional FPA arrays such that the interference leakage of satellites can be more effectively suppressed \cite{zhu2023MAMag}. The superiority of MA arrays over FPA arrays in terms of flexible beamforming has been validated in existing literatures \cite{zhu2023MAarray,ma2024multi}. It was shown in \cite{zhu2023MAarray} that the full array gain over the desired direction and interference nulling over multiple undesired directions can be simultaneously achieved by MA arrays. Moreover, by alternately optimizing the APV and AWV, the MA array can attain considerable performance gain in multi-beam forming compared to conventional FPA arrays \cite{ma2024multi}.

Recently, the MA-aided wireless communications have been widely investigated in terrestrial systems. For example, based on the field-response channel model or spatial correlation channel model, it was revealed that the local movement of MAs at the transmitter/receiver can attain considerable improvement in the received signal power for both narrow-band and wideband systems \cite{zhu2022MAmodel,wong2020limit,mei2024movable,zhu2024wideband}. The MA-enabled multiple-input multiple-output (MIMO) systems were investigated in \cite{ma2022MAmimo} and \cite{chen2023joint} based on instantaneous channel state information (CSI) and statistical CSI, respectively, where the optimization of multiple MAs' positions can significantly increase the MIMO capacity. The MA-aided multiuser communications have also been studied under different system setups \cite{zhu2023MAmultiuser,xiao2023multiuser,wu2023movable,Wong2023opport,hu2024power,qin2024antenna,cheng2023sum,Yang2024movable,wang2024multiuser}. Therein, the MA position optimization can reconfigure the wireless channels between the BS and users such that the multiuser interference is efficiently mitigated and the spatial multiplexing performance can be improved. Moreover, the studies in \cite{hu2024secure,cheng2024secure,tang2024secure} demonstrated the great potential of MAs in enhancing the physical-layer security of wireless communication systems. The channel estimation for MA systems was explored in \cite{ma2023MAestimation,xiao2023channel} to recovery the field-response information between the transmitter and receiver regions based on compressed sensing. Besides, a new architecture of six-dimensional MA (6DMA) was proposed in \cite{shao20246DMA,shao2024discrete} by jointly designing the three-dimensional (3D) position and 3D rotation of antenna surfaces at the BS based on the users' spatial distribution/statistical CSI.

Despite the above works on terrestrial communication systems \cite{zhu2023MAarray,ma2024multi,zhu2022MAmodel,wong2020limit,mei2024movable,zhu2024wideband,ma2022MAmimo,chen2023joint,zhu2023MAmultiuser,xiao2023multiuser,wu2023movable,Wong2023opport,hu2024power,qin2024antenna,cheng2023sum,Yang2024movable,wang2024multiuser,hu2024secure,cheng2024secure,tang2024secure,ma2023MAestimation,xiao2023channel,shao20246DMA,shao2024discrete}, MA-aided satellite communication has not been investigated, which motivates this paper to exploit the newly introduced DoF of MA arrays in enhancing the dynamic beam coverage and interference mitigation for LEO satellite networks. The main contributions of this paper are summarized as follows:

\begin{itemize}
	\item We consider the MA array-aided beam coverage of a typical satellite in an LEO constellation. Given the satellite orbit and coverage requirement within a specific time interval, the APV and AWV of the satellite-mounted MA array are jointly optimized over time to minimize the average signal leakage power to the interference area of the satellite, subject to the constraints of the minimum beamforming gain over the coverage area, the feasible moving region of MAs, the minimum inter-MA spacing, the maximum moving speed of MAs, and the constant modulus of AWV.
	\item Since the formulated problem involves the continuous-time decision process of the APV and AWV, we first transform it into a more tractable discrete-time optimization problem by discretizing the time interval and the angular coordinate of terrestrial areas. Then, an alternating optimization (AO)-based algorithm is developed by iteratively optimizing the APV and AWV over all discrete time slots. In particular, the successive convex approximation (SCA) technique is utilized to convexify the APV/AWV optimization subproblem for obtaining locally optimal solutions during the iterations.
	\item Considering the high control overhead and energy consumption for continuous movement of MAs in real time, we propose a low-complexity scheme by using an optimized common APV over all time slots. As such, the positions of MAs can be reconfigured at the first time slot based on the coverage requirement, while further movement is dispensed at the subsequential time slots until the coverage requirement of the satellite changes. The corresponding problem for joint APV and AWV optimization can also be solved by the proposed AO-based algorithm, where the computational complexity of APV optimization is significantly reduced. 
	\item Finally, simulation results are presented to validate the efficacy of MA array-enhanced dynamic beam coverage for LEO satellite communications. Compared to FPA arrays such as the conventional uniform planar array (UPA), the proposed MA array-aided beamforming schemes can significantly decrease the interference leakage and increase the average signal-to-leakage ratio (SLR) of satellite-ground links. Moreover, the low-complexity MA scheme can achieve a performance comparable to the continuous-movement MA scheme in terms of interference mitigation, subject to the same coverage performance requirement. 
\end{itemize}


The rest of this paper is organized as follows. In Section II, we introduce the system model and formulate the MA-aided satellite beamforming problem. In Section III, we present the proposed solutions for the formulated optimization problem. Then, we show the simulation results in Section IV and this paper is finally concluded in Section V.

\textit{Notation}: $a$, $\mathbf{a}$, $\mathbf{A}$, and $\mathcal{A}$ denote a scalar, a vector, a matrix, and a set, respectively. $(\cdot)^{\rm{T}}$ and $(\cdot)^{\rm{H}}$ denote transpose and conjugate transpose, respectively. $\mathcal{A} \setminus \mathcal{B}$ and $\mathcal{A} \cap \mathcal{B}$ represent the subtraction set and intersection set of $\mathcal{A}$ and $\mathcal{B}$, respectively. $\mathbb{R}^{M \times N}$ and $\mathbb{C}^{M \times N}$ represent the sets of real and complex matrices/vectors of dimension $M \times N$, respectively. $|\cdot|$ and $\angle(\cdot)$ denote the amplitude and the phase of a complex number or complex vector, respectively. $\|\cdot\|_{2}$ denotes the 2-norm of a vector. $\dif(\cdot)$ denotes the differential of a variable/function. $\mathbf{1}_{M \times 1}$ denotes an $M$-dimensional column vector with all elements equal to 1. $\otimes$ denotes the Kronecker product. $f(x|y)$ represents a function with respect to (w.r.t.) $x$ determined by parameter $y$.

\section{System Model}
\subsection{Satellite Orbit Model}
\begin{figure}[t]
	\begin{center}
		\subfigure[LEO satellite constellation.]{\includegraphics[width=8.8 cm]{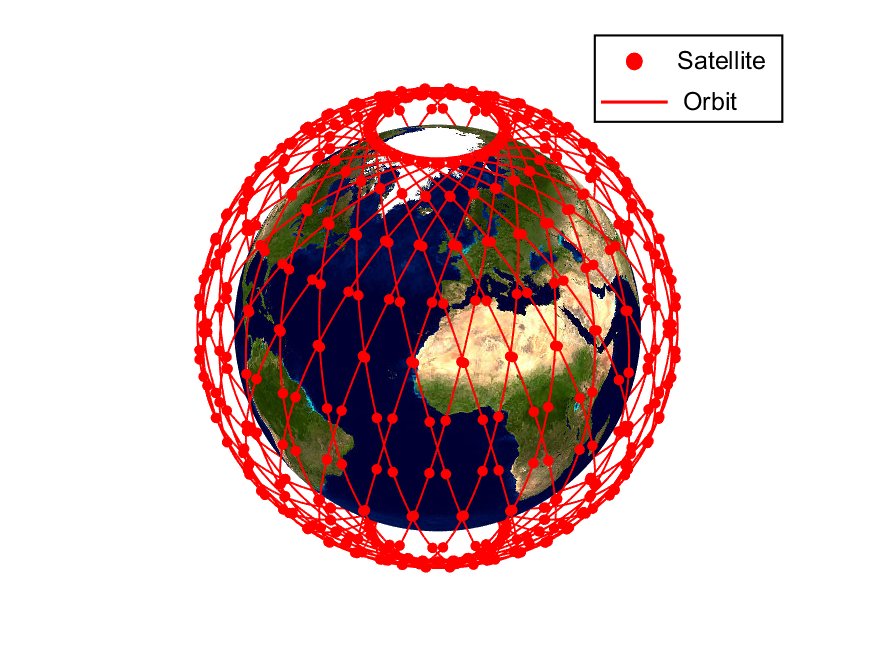} \label{fig:constellation}}
		\subfigure[Geocentric coordinate system.]{\includegraphics[width=7 cm]{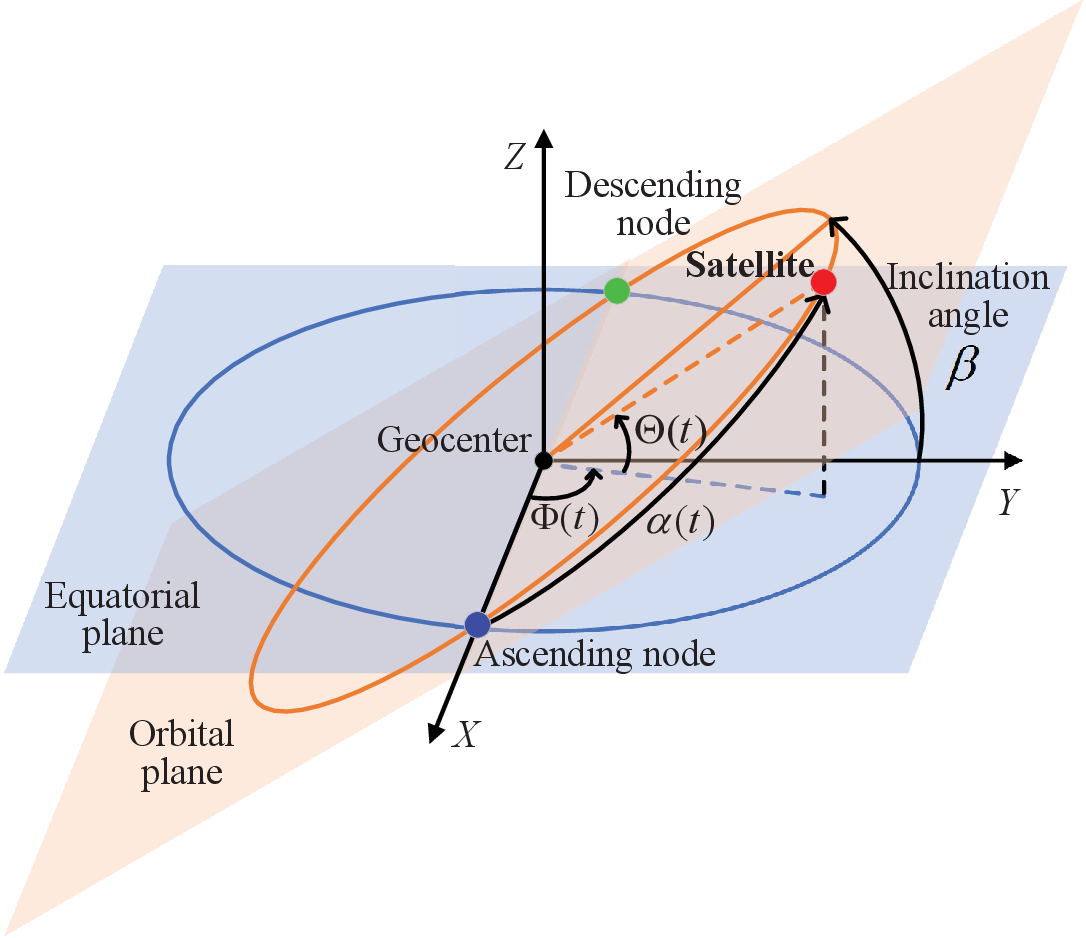} \label{fig:SphereCoor}}
		\caption{Illustration of the considered LEO satellite constellation and the geocentric coordinate system.}
	\end{center}
\end{figure}

In this paper, we consider a typical Walker Delta LEO constellation shown in Fig. \ref{fig:constellation}\footnote{The MA array-aided beam coverage solution proposed in this paper can be applied to any other constellations with given orbit and coverage requirement of each satellite.}, where the number of orbital planes and the number of satellites in each orbital plane are denoted by $K_{\mathrm{o}}$ and $K_{\mathrm{s}}$, respectively. The orbital inclination angle is denoted by $\beta$, which represents the angle between an orbital plane and the equatorial plane as shown in Fig. \ref{fig:constellation}. Without loss of generality, we suppose that the earth is a regular sphere with radius $R_{\mathrm{e}}$. The altitude of the satellite orbit relative to the earth surface is denoted by $H_{\mathrm{s}}$. Then, the orbital period, i.e., the time that a satellite takes to complete one orbit around the earth, is obtained as $T_{\mathrm{s}}=2\pi \sqrt{\frac{(R_{\mathrm{e}}+H_{\mathrm{s}})^{3}}{G_{\mathrm{e}}M_{\mathrm{e}}}}$, where $G_{\mathrm{e}}$ and $M_{\mathrm{e}}$ are the constant of gravitation and the mass of the earth, respectively.

Since the orbits of satellites are deterministic, the locations of satellites for any given time are known {\it a priori}. In the 3D space, we consider a geocentric spherical coordinate system (GSCS) with the equatorial plane being the reference plane as shown in Fig. \ref{fig:SphereCoor}. For any point in the 3D space, the angle from the reference plane to the direction of this point is defined as the elevation angle $\Theta$. The direction from south to north is defined as the positive sense of the elevation angle, which is thus ranging from $-\pi/2$ to $\pi/2$. The direction towards the intersection between the equator and the prime meridian is defined as the azimuth reference direction. The azimuth angle $\Phi$ refers to the angle from the azimuth reference direction to the direction of the projection of a point on the reference plane. Define the direction from west to east as the positive sense of the azimuth angle. Thus, the range of the azimuth angle is from $-\pi$ to $\pi$.

Without loss of generality, we consider a typical satellite in an orbital plane, while the other satellites beam coverage can be designed in a similar manner. As shown in Fig \ref{fig:SphereCoor}, the geocentric angle between the directions of the ascending node and the satellite at time $t$ is given by $\alpha(t) = 2\pi t /T_{\mathrm{s}} + \alpha_{0}$ (in radian), where $\alpha_{0}$ denotes the initial angle at $t=0$. Then, the coordinates of the typical satellite in the GSCS can be expressed as \cite{montenbruck2002satellite} 
\begin{equation} \left\{
	\begin{aligned}
		&R_{\mathrm{s}}(t) = R_{\mathrm{e}} + H_{\mathrm{s}},\\
		&\Theta_{\mathrm{s}}(t) = \arcsin\left[\sin \beta \sin \alpha(t)  \right] \in [-\pi/2, \pi/2],\\
		&\Phi_{\mathrm{s}}(t) = \arctan\left[\cos \beta \tan \alpha(t) \right] \in (-\pi, \pi].
	\end{aligned}\right.
\end{equation}
The corresponding coordinates in the 3D geocentric Cartesian coordinate system (GCCS) is thus given by {\footnotesize$\left[R_{\mathrm{s}}(t) \cos \Theta_{\mathrm{s}}(t) \cos \Phi_{\mathrm{s}}(t), R_{\mathrm{s}}(t) \cos \Theta_{\mathrm{s}}(t) \sin \Phi_{\mathrm{s}}(t), R_{\mathrm{s}}(t) \sin \Theta_{\mathrm{s}}(t) \right]^{\mathrm{T}}$}. It is worth noting that the $K_{\mathrm{s}}$ satellites in the same orbital plane has the same trajectory. In practice, they can simultaneously switch their coverage areas, with each subsequent satellite taking over the coverage area of the previous satellite in this orbital plane. Thus, the beam coverage solution for the typical satellite can also be applied to its subsequent satellites in this orbital plane. As such, we only need to focus on the beam coverage problem within one typical time interval $(0,T]$, with $T=T_{\mathrm{s}}/K_{\mathrm{s}}$.\footnote{For LEO constellations, $T$ is much smaller than the rotation period of the earth, and thus the impact of earth rotation is neglected throughout this paper.}

\begin{figure}[t]
	\begin{center}
		\includegraphics[width=8.8 cm]{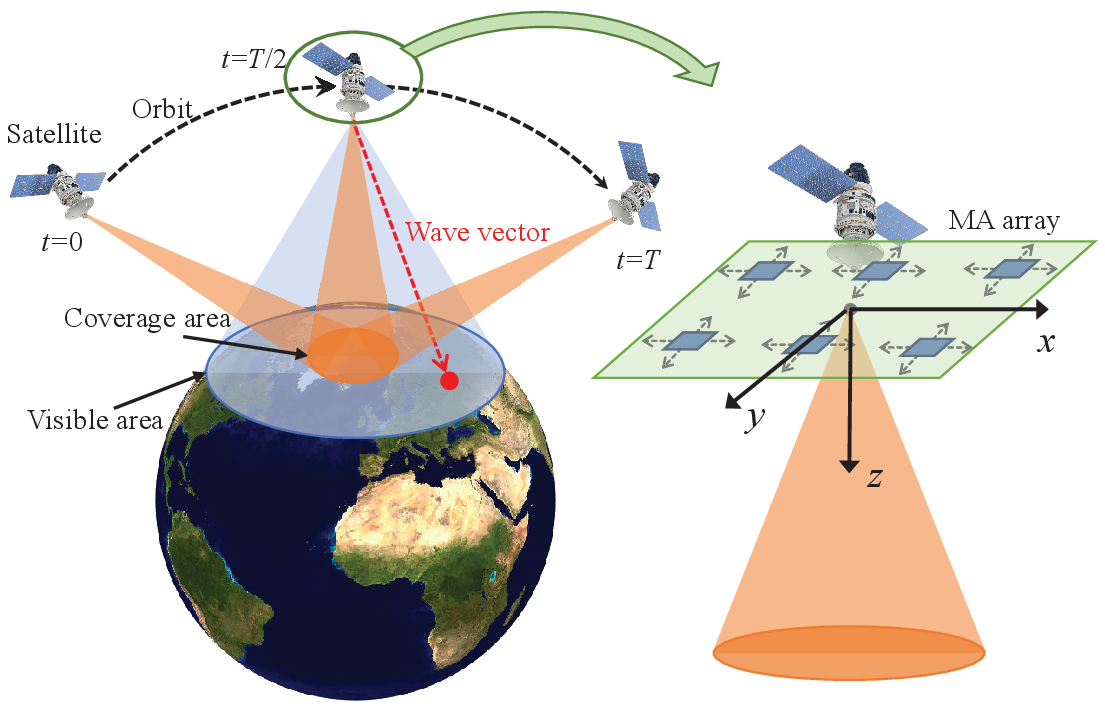}
		\caption{Illustration of the coverage area and the SCCS.}
		\label{fig:satellite_cov}
	\end{center}
\end{figure}

\subsection{Beam Coverage Model}
Given the GSCS coordinates of a point on the earth surface, i.e., $[R_{\mathrm{e}},\Theta,\Phi]^{\mathrm{T}}$, the wave vector (from the point on the earth to the satellite) in the GCCS at time $t$ is given by 
\begin{equation}\label{eq_wavevec}
	\mathbf{k}_{\mathrm{e}}(\Theta, \Phi, t)=\frac{2 \pi}{\lambda} \frac{\bar{\mathbf{k}}_{\mathrm{e}}(\Theta, \Phi, t)}{\left\|\bar{\mathbf{k}}_{\mathrm{e}}(\Theta, \Phi, t)\right\|_2},
\end{equation}
with $\bar{\mathbf{k}}_{\mathrm{e}}(\Theta, \Phi, t)$ given by \eqref{eq_wavevec2} at the top of next page.
\begin{figure*}[!t]
	\centering 
	\begin{equation}\label{eq_wavevec2}
		\bar{\mathbf{k}}_{\mathrm{e}}(\Theta, \Phi, t)=\left[\begin{array}{l}
			R_{\mathrm{e}} \cos \Theta \cos \Phi \\
			R_{\mathrm{e}} \cos \Theta \sin \Phi \\
			R_{\mathrm{e}} \sin \Theta
		\end{array}\right]-\left[\begin{array}{l}
			\left(R_{\mathrm{e}}+H_{\mathrm{s}}\right) \cos \Theta_{\mathrm{s}}(t) \cos \Phi_{\mathrm{s}}(t) \\
			\left(R_{\mathrm{e}}+H_{\mathrm{s}}\right) \cos \Theta_{\mathrm{s}}(t) \sin \Phi_{\mathrm{s}}(t) \\
			\left(R_{\mathrm{e}}+H_{\mathrm{s}}\right) \sin \Theta_{\mathrm{s}}(t)
		\end{array}\right],
	\end{equation}
	\begin{equation}\label{eq_transform}
		\mathbf{T}_{\mathrm{s}}(t)=\left[\begin{array}{ccc}
			-\sin \beta \sin \Theta_{\mathrm{s}}(t)-\cos \beta \cos \Theta_{\mathrm{s}}(t) \sin \Phi_{\mathrm{s}}(t) & 0 & -\cos \Theta_{\mathrm{s}}(t) \cos \Phi_{\mathrm{s}}(t) \\
			\cos \beta \cos \Theta_{\mathrm{s}}(t) \cos \Phi_{\mathrm{s}}(t) & \sin \beta & -\cos \Theta_{\mathrm{s}}(t) \sin \Phi_{\mathrm{s}}(t) \\
			\sin \beta \cos \Theta_{\mathrm{s}}(t) \cos \Phi_{\mathrm{s}}(t) & -\cos \beta & -\sin \Theta_{\mathrm{s}}(t)
		\end{array}\right].
	\end{equation}
	\vspace*{8pt}
	\hrulefill
\end{figure*}
To facilitate the beam coverage modeling of the satellite, we establish a satellite-centric Cartesian coordinate system (SCCS) as shown in Fig. \ref{fig:satellite_cov}. In particular, axis $x$ is defined as the satellite's moving direction, axis $y$ is orthogonal to the orbital plane, and axis $z$ directs to the geocenter, respectively. According to basic geometry, the coordinate transform matrix from the SCCS to the GCCS is given by \eqref{eq_transform} shown at the top of next page. Thus, the coordinates of the wave vector in the SCCS is given by 
\begin{equation}\label{eq_wavevec_sat}
	\mathbf{k}(\Theta, \Phi, t) = \mathbf{T}_{\mathrm{s}}(t)^{\mathrm{T}} \mathbf{k}_{\mathrm{e}}(\Theta, \Phi, t).
\end{equation}

The satellite is equipped with an MA array with $N$ elements, where the coordinates of the $n$-th MA in the SCCS at time $t$ is denoted by $\bar{\mathbf{q}}_{n}(t) = [x_{n}(t), y_{n}(t), z_{n}(t)]^{\mathrm{T}}$, $1 \leq n \leq N$, $t \in (0,T]$. Without loss of generality, we assume in this paper that each MA can only move within a two-dimensional (2D) area $\mathcal{C}$ in the $x$-$y$ plane and thus we always have $\bar{\mathbf{q}}_{n}(t)=[\mathbf{q}_{n}(t)^{\mathrm{T}}, 0]^{\mathrm{T}}$ and $\mathbf{q}_{n}(t) \in \mathcal{C} \subseteq \mathbb{R}^{2 \times 1}$. Denoting the collective APV as $\mathbf{q}(t)=[\mathbf{q}_{1}(t)^{\mathrm{T}}, \mathbf{q}_{2}(t)^{\mathrm{T}}, \dots, \mathbf{q}_{N}(t)^{\mathrm{T}}]^{\mathrm{T}}$, which determines the geometry of the MA array. The steering vector of the MA array is thus given by
\begin{equation}\label{eq_steer_vec}
	\mathbf{a}(\mathbf{k}(\Theta, \Phi, t), \mathbf{q}(t))=\left[e^{j \mathbf{k}(\Theta, \Phi, t)^{\mathrm{T}} \bar{\mathbf{q}}_n(t)}\right]_{1 \leq n \leq N}^{\mathrm{T}},
\end{equation}
which represents the phase shifts at all MA elements for the plane wave with wave vector $\mathbf{k}(\Theta, \Phi, t)$. 

Since all the points on the earth surface have the same radial distance $R_{\mathrm{e}}$ in the GSCS, for notation simplicity, we use angular coordinate $(\Theta,\Phi)$ to represent the position of a point on the earth surface. Denote the coverage area on the earth as $\mathcal{A}_{\mathrm{e}}$, which is a constant set in the GSCS. For any given location of the satellite, it can only establish line-of-sight (LoS) links with a subarea on the earth as shown in Fig. \ref{fig:satellite_cov}, while the remaining area is blocked by the earth. Thus, we can define a visible area of the satellite to represent the set of points on the earth surface that can receive signals from the satellite. According to basic geometry, the visible area on the earth at time $t$ can be obtained as
\begin{equation}\label{eq_visible_area}
	\mathcal{A}_{\mathrm{s}}(t)=\left\{(\Theta, \Phi) {\Big\vert}  \left\|\bar{\mathbf{k}}_{\mathrm{e}}(\Theta, \Phi, t)\right\|_2^2 \leq\left(R_{\mathrm{e}}+H_{\mathrm{s}}\right)^2-R_{\mathrm{e}}^2\right\},
\end{equation}
where $\left\|\bar{\mathbf{k}}_{\mathrm{e}}(\Theta, \Phi, t)\right\|_2$ represents the distance between the satellite and the point $(\Theta, \Phi)$ on the earth surface and $\left\|\bar{\mathbf{k}}_{\mathrm{e}}(\Theta, \Phi, t)\right\|_2^2 = \left(R_{\mathrm{e}}+H_{\mathrm{s}}\right)^2-R_{\mathrm{e}}^2$ indicates that vector $\bar{\mathbf{k}}_{\mathrm{e}}(\Theta, \Phi, t)$ is tangent to the earth surface. Then, the interference area of the satellite at time $t$ is defined as\footnote{The interference area can also adopt other definitions based on the distributions of terrestrial terminals, which do not impact the mathematical formulation of the considered satellite beam coverage problem.}
\begin{equation}\label{eq_interf_area}
	\mathcal{A}_{\mathrm{i}}(t) = \mathcal{A}_{\mathrm{s}}(t)\setminus\mathcal{A}_{\mathrm{e}}.
\end{equation}

Denoting $\gamma$ as the path loss exponent and $\rho_{0}$ as the path loss for the reference distance of 1 meter (m), the path loss between the satellite and point $(\Theta,\Phi)$ on the earth surface is given by $\rho(\Theta, \Phi, t) = \rho_{0} \left\|\bar{\mathbf{k}}_{\mathrm{e}}(\Theta, \Phi, t)\right\|_{2}^{-\gamma}$. Then, the LoS channel vector between the MA array at the satellite and point $(\Theta,\Phi)$ on the earth surface is obtained as 
\begin{equation}
	\sqrt{\rho(\Theta, \Phi, t)}e^{j\frac{2\pi}{\lambda}\left\|\bar{\mathbf{k}}_{\mathrm{e}}(\Theta, \Phi, t)\right\|_{2}}\mathbf{a}(\mathbf{k}(\Theta, \Phi, t), \mathbf{q}(t)).
\end{equation}
Denoting $\mathbf{w}(t) =[w_{1}(t),w_{2}(t),\dots,w_{N}(t)]^{\mathrm{T}} \in \mathbb{C}^{N \times 1}$ as the AWV at time $t$, the effective channel gain between the satellite and point $(\Theta,\Phi)$ on the earth surface can be expressed as
\begin{equation}\label{eq_eff_channel_gain}
	\begin{aligned}
		&h(\Theta, \Phi, \mathbf{q}(t), \mathbf{w}(t), t)\\
		=&\rho(\Theta, \Phi, t)\left|\mathbf{a}(\mathbf{k}(\Theta, \Phi, t), \mathbf{q}(t))^{\mathrm{H}} \mathbf{w}(t) \right|^{2}, ~t \in (0,T].
	\end{aligned}
\end{equation}

\subsection{Problem Formulation}
In practice, to improve the communication performance of terrestrial terminals, it is desired to increase the beamforming gain over the coverage area and decrease the signal leakage to the interference area simultaneously. To this end, the average beamforming gain over the coverage area at time $t$ is defined as 
\begin{equation}\label{eq_beamgain_ave}
	\begin{aligned}
	&G(\mathbf{q}(t), \mathbf{w}(t),t) \\
	= &\frac{\iint \limits_{(\Theta,\Phi) \in \mathcal{A}_{\mathrm{e}}} h(\Theta, \Phi, \mathbf{q}(t), \mathbf{w}(t), t) \dif \Theta \dif \Phi}
	{\iint \limits_{(\Theta,\Phi) \in \mathcal{A}_{\mathrm{e}}} \rho(\Theta, \Phi, t) \dif \Theta \dif \Phi}, ~t \in (0,T],
	\end{aligned}
\end{equation}
where the path loss over the entire coverage area is normalized. Moreover, the average signal leakage power to the interference area within time interval $(0,T]$ is given by
\begin{equation}\label{eq_interfer_ave}
	\begin{aligned}
		&I(\mathbf{q}(t), \mathbf{w}(t)) \\
		= &\frac{1}{T} \int \limits_{t \in (0,T]}
		\frac{\iint \limits_{(\Theta,\Phi) \in \mathcal{A}_{\mathrm{i}}(t)} h(\Theta, \Phi, \mathbf{q}(t), \mathbf{w}(t), t) \dif \Theta \dif \Phi}
		{\iint \limits_{(\Theta,\Phi) \in \mathcal{A}_{\mathrm{i}}(t)} \rho(\Theta, \Phi, t) \dif \Theta \dif \Phi} \dif t,
	\end{aligned}
\end{equation}
where the path loss over the entire interference area is first normalized and the leakage power is then averaged over time interval $(0,T]$. 

We aim to minimize the interference leakage power while guaranteeing the average beamforming gain over the coverage area at any time $t \in (0,T]$. The joint APV and AWV optimization problem for satellite beam coverage is thus formulated as
\begin{subequations}\label{eq_problem}
	\begin{align}
		&\mathop{\min}\limits_{\{\mathbf{q}(t), \mathbf{w}(t)\}_{t\in (0,T]}}~
		I(\mathbf{q}(t), \mathbf{w}(t)) \label{eq_problem_a}\\
		&~~~~~~~\mathrm{s.t.}~~  G(\mathbf{q}(t), \mathbf{w}(t), t) \geq \eta, ~t \in (0,T], \label{eq_problem_b}\\
		&~~~~~~~~~~~~~\mathbf{q}_n(t) \in \mathcal{C},~ t \in(0, T],1 \leq n \leq N, \label{eq_problem_c}\\
		&~~~~~~~~~~~~~\left\|\mathbf{q}_n(t)-\mathbf{q}_{\hat{n}}(t)\right\|_2 \geq d_{\min },  t \in(0, T], n \neq \hat{n} , \label{eq_problem_d}\\
		&~~~~~~~~~~~~~\left\|\frac{\dif \mathbf{q}_n(t)}{\dif t}\right\|_2 \leq v_{\max },~ t \in(0, T], 1 \leq n \leq N, \label{eq_problem_e}\\
		&~~~~~~~~~~~~~\left|w_{n}(t)\right|=\frac{1}{\sqrt{N}},~t \in(0, T], 1 \leq n \leq N,  \label{eq_problem_f}
	\end{align}
\end{subequations}
where constraint \eqref{eq_problem_b} ensures the average beamforming gain over the coverage area with $\eta$ being the minimum threshold; constraint \eqref{eq_problem_c} confines the antenna moving region; constraint \eqref{eq_problem_d} guarantees that the inter-antenna spacing should be no less than the minimum value $d_{\min}$; constraint \eqref{eq_problem_e} restrains the maximum moving speed of each MA as $v_{\max}$; and constraint \eqref{eq_problem_f} is the constant-modulus constraint for the AWV under the analog beamforming structure. Problem \eqref{eq_problem} is a non-convex optimization problem. On one hand, the optimization variables involve the positions and weights of $N$ MAs over the continuous time interval $(0,T]$, which have a high dimension. On the other hand, the APVs and AWVs are highly coupled at different time $t$. Thus, it is challenging to derive the globally optimal solution for problem \eqref{eq_problem}. In the following, we will propose an AO-based algorithm to obtain a suboptimal solution.

\section{Proposed Solution}
In this section, we first transform problem \eqref{eq_problem} into a tractable problem by implementing discretization on time and angle. Then, the AO over the APV and AWV is applied by employing the SCA technique for each iteration. Finally, an alternative scheme is proposed as a low-complexity solution for antenna movement.

\subsection{Problem Transformation}
Problem \eqref{eq_problem} requires to solve $\mathbf{q}(t)$ and $\mathbf{w}(t)$ over continuous time interval $(0,T]$, which has an infinite dimension. To facilitate the problem solving, we equally divide the interval $(0,T]$ into $M$ time slots, i.e., $\{(\frac{(m-1)T}{M},\frac{mT}{M}]\}_{1\leq m \leq M}$. The number of time slots is chosen to be sufficiently large such that $\Theta_{\mathrm{s}}(t)$ and $\Phi_{\mathrm{s}}(t)$ can be approximately considered as unchanged within each time slot. Then, we can use the snapshot at time $t_{m}=\frac{(m-1/2)T}{M}$ to represent the average performance of the $m$-th time slot, $1 \leq m \leq M$. As such, $\mathbf{q}(t)$ and $\mathbf{w}(t)$ can be replaced by 
\begin{subequations}\label{eq_APV_AWV_discrete}
	\begin{align}
		&\mathbf{q} \triangleq \left[\mathbf{q}[1]^{\mathrm{T}},\mathbf{q}[2]^{\mathrm{T}},\dots,\mathbf{q}[M]^{\mathrm{T}} \right]^{\mathrm{T}} \in \mathbb{R}^{2MN \times 1}, \\
		&\mathbf{w} \triangleq \left[\mathbf{w}[1]^{\mathrm{T}},\mathbf{w}[2]^{\mathrm{T}},\dots,\mathbf{w}[M]^{\mathrm{T}} \right]^{\mathrm{T}} \in \mathbb{C}^{MN \times 1}, 
	\end{align}
\end{subequations}
with $\mathbf{q}[m] \triangleq \mathbf{q}(t_m)$ and $\mathbf{w}[m] \triangleq \mathbf{w}(t_m)$, $1 \leq m \leq M$.

Moreover, to approximate the integral w.r.t. $\Theta$ and $\Phi$, we equally divide interval $[-\pi/2, \pi/2] \times (-\pi, \pi]$ into $L_{\mathrm{e}} \times L_{\mathrm{a}}$ grids, with the set of their centers given by
\begin{equation}\label{eq_grid_center}
	\begin{aligned}
		&\mathcal{A}_{\mathrm{d}} = {\Big\{}(\Theta_{l_{\mathrm{e}}}, \Phi_{l_{\mathrm{a}}}) {\Big\vert} \Theta_{l_{\mathrm{e}}} = -\frac{\pi}{2} + \frac{(l_{\mathrm{e}}-1/2)\pi}{L_{\mathrm{e}}}, \\
		&~~\Phi_{l_{\mathrm{a}}} = -\pi+\frac{(2l_{\mathrm{a}}-1)\pi}{L_{\mathrm{a}}}, 1 \leq l_{\mathrm{e}} \leq L_{\mathrm{e}}, 1 \leq l_{\mathrm{a}} \leq L_{\mathrm{a}}{\Big\}}. 
	\end{aligned}
\end{equation}
For notation simplicity, the set of discretized grids in the coverage area and that in the interference area are respectively denoted as
\begin{subequations}\label{eq_cover_area}
	\begin{align}
		&\bar{\mathcal{A}}_{\mathrm{e}} = \mathcal{A}_{\mathrm{d}} \cap \mathcal{A}_{\mathrm{e}} \triangleq\left\{\left(\Theta_l, \Phi_l\right), 1 \leq l \leq L_0\right\}, \\
		&\bar{\mathcal{A}}_{\mathrm{i}}^{m} = \mathcal{A}_{\mathrm{d}} \cap \mathcal{A}_{\mathrm{i}}\left(t_m\right) \triangleq\left\{\left(\Theta_{m, l}, \Phi_{m, l}\right), 1 \leq l \leq L_m\right\}, 
	\end{align}
\end{subequations}
where $L_{0}$ denotes the total number of discretized grids in the coverage area and $L_{m}$ denotes that in the interference area at time $t_{m}$, $1 \leq m \leq M$, respectively.

Based on the above discretization, the average beamforming gain of the coverage area in \eqref{eq_beamgain_ave} at the $m$-th time slot can be approximated by
\begin{equation}\label{eq_beamgain_ave2}
	\begin{aligned}
		&G_{m}(\mathbf{q}[m], \mathbf{w}[m]) = \sum\limits_{l=1}^{L_0} g_{m, l}^{\mathrm{cov}}\left|\mathbf{a}\left(\mathbf{k}_{m, l}^{\mathrm{cov}}, \mathbf{q}[m]\right)^{\mathrm{H}} \mathbf{w}[m]\right|^2,
	\end{aligned}
\end{equation}
with $\mathbf{q}[m] \triangleq \mathbf{q}(t_{m})$, $\mathbf{w}[m] \triangleq \mathbf{w}(t_{m})$, $\mathbf{k}_{m, l}^{\mathrm{cov}} \triangleq \mathbf{k}(\Theta_{l}, \Phi_{l}, t_{m})$, and
$$g_{m, l}^{\mathrm{cov}} \triangleq \frac{\rho(\Theta_{l},\Phi_{l},t_{m})}{\sum \limits_{j=1}^{L_{0}} \rho(\Theta_{j},\Phi_{j},t_{m})}, ~1 \leq m \leq M, 1 \leq l \leq L_{0}.$$
Similarly, the average signal leakage power to the interference area in \eqref{eq_interfer_ave} is approximated by
\begin{equation}\label{eq_interfer_ave2}
	\begin{aligned}
		&I(\mathbf{q}, \mathbf{w}) = \sum \limits_{m=1}^{M} I_{m}(\mathbf{q}[m], \mathbf{w}[m]) \\
		= & \sum \limits_{m=1}^{M}\sum \limits_{l=1}^{L_m} g_{m, l}^{\mathrm{int}}\left|\mathbf{a}\left(\mathbf{k}_{m, l}^{\mathrm{int}}, \mathbf{q}[m]\right)^{\mathrm{H}} \mathbf{w}[m]\right|^2,
	\end{aligned}
\end{equation}
with $\mathbf{k}_{m, l}^{\mathrm{int}} \triangleq \mathbf{k}(\Theta_{m,l}, \Phi_{m,l}, t_{m})$, and
$$g_{m, l}^{\mathrm{int}} \triangleq \frac{\rho(\Theta_{m,l},\Phi_{m,l},t_{m})}{\sum \limits_{j=1}^{L_{m}} \rho(\Theta_{m,j},\Phi_{m,j},t_{m})}, ~ 1 \leq m \leq M, 1 \leq l \leq L_{m}.$$

Since the interval of each time slot is sufficiently small, the position of each antenna can be regarded as unchanged within each time slot and the moving speed of the $n$-th MA at the $m$-th time slot can be approximated by $\frac{M}{T}\|\mathbf{q}_{n}[m+1] - \mathbf{q}_{n}[m]\|_{2}$. As such, problem \eqref{eq_problem} can be converted into the discrete form as follows:
\begin{subequations}\label{eq_problem_dis}
	\begin{align}
		\mathop{\min}\limits_{\mathbf{q}, \mathbf{w}}~
		&I(\mathbf{q}, \mathbf{w}) \label{eq_problem_dis_a}\\
		\mathrm{s.t.}~~ & G_{m}(\mathbf{q}[m], \mathbf{w}[m]) \geq \eta,~ 1 \leq m \leq M, \label{eq_problem_dis_b}\\
		&\mathbf{q}_n[m] \in \mathcal{C}, ~ 1 \leq m \leq M, 1 \leq n \leq N,  \label{eq_problem_dis_c}\\
		&\left\|\mathbf{q}_n[m]-\mathbf{q}_{\hat{n}}[m]\right\|_2 \geq d_{\min },~ 1 \leq m \leq M, n \neq \hat{n}, \label{eq_problem_dis_d}\\
		&\begin{aligned}&\left\|\mathbf{q}_{n}[m+1] - \mathbf{q}_{n}[m] \right\|_2 \leq \frac{v_{\max}T}{M},\\
			&~~~~~~~~~~~~~~~~~~~~~~ 1 \leq m \leq M-1, 1 \leq n \leq N, \label{eq_problem_dis_e} \end{aligned}\\
		&\left|w_{n}[m]\right|=\frac{1}{\sqrt{N}},~1 \leq m \leq M, 1 \leq n \leq N. \label{eq_problem_dis_f}
	\end{align}
\end{subequations}
Note that problem \eqref{eq_problem_dis} is still non-convex w.r.t. both $\mathbf{q}$ and $\mathbf{w}$. In the following, we alternately optimize the APV $\mathbf{q}$ and the AWV $\mathbf{w}$ with the other one being fixed and obtain a suboptimal solution for problem \eqref{eq_problem_dis}.

\subsection{Optimization of APV}
Given AWV $\mathbf{w}^{(i-1)}$ obtained in the $(i-1)$-th iteration of the AO, the APV optimization in the $i$-th iteration is to solve the following subproblem:
\begin{subequations}\label{eq_problem_APV}
	\begin{align}
		\mathop{\min}\limits_{\mathbf{q}}~
		&I(\mathbf{q}, \mathbf{w}^{(i-1)}) \label{eq_problem_APV_a}\\
		\mathrm{s.t.}~~ & G_{m}(\mathbf{q}[m], \mathbf{w}^{(i-1)}[m]) \geq \eta,~ 1 \leq m \leq M, \label{eq_problem_APV_b}\\
		&\mathbf{q}_n[m] \in \mathcal{C}, 1 \leq m \leq M,~ 1 \leq n \leq N,  \label{eq_problem_APV_c}\\
		&\left\|\mathbf{q}_n[m]-\mathbf{q}_{\hat{n}}[m]\right\|_2 \geq d_{\min },~ 1 \leq m \leq M, n \neq \hat{n}, \label{eq_problem_APV_d}\\
		&\begin{aligned}&\left\|\mathbf{q}_{n}[m+1] - \mathbf{q}_{n}[m] \right\|_2 \leq \frac{v_{\max}T}{M},\\
			&~~~~~~~~~~~~~~~~~~~~~~ 1 \leq m \leq M-1, 1 \leq n \leq N. \label{eq_problem_APV_e} \end{aligned}
	\end{align}
\end{subequations}
The objective function in \eqref{eq_problem_APV_a} and constraints \eqref{eq_problem_APV_b} and \eqref{eq_problem_APV_d} are not convex. To address this issue, we construct surrogate functions for $I(\mathbf{q}, \mathbf{w}^{(i-1)})$ and $G_{m}(\mathbf{q}[m], \mathbf{w}^{(i-1)}[m])$ at local point $\mathbf{q}^{(i-1)}$ such that problem \eqref{eq_problem_APV} can be converted into a convex optimization problem w.r.t. $\mathbf{q}$. 

To this end, for any real constant vector $\mathbf{k}$ and scalar $\varphi$, we derive the second-order Taylor expansion of function $\hat{f}(\mathbf{q} \vert \mathbf{k},\varphi) = \cos(\mathbf{k}^{\mathrm{T}}\mathbf{q}+\varphi)$ w.r.t. $\mathbf{q}$ at any given point $\mathbf{q}_{0}$ as 
\begin{equation}\label{eq_Taylor_cos}
	\begin{aligned}	
		&\cos(\mathbf{k}^{\mathrm{T}}\mathbf{q}_{0}+\varphi) - \sin(\mathbf{k}^{\mathrm{T}}\mathbf{q}_{0}+\varphi)\mathbf{k}^{\mathrm{T}}(\mathbf{q} - \mathbf{q}_{0})\\
		& ~~~~~~~~~~~~~~~~~~~-\frac{1}{2} \cos(\mathbf{k}^{\mathrm{T}}\mathbf{q}_{0}+\varphi) \left(\mathbf{k}^{\mathrm{T}}(\mathbf{q} - \mathbf{q}_{0})\right)^{2}.
	\end{aligned}
\end{equation}
Given the fact of $-1 \leq \cos(\mathbf{k}^{\mathrm{T}}\mathbf{q}_{0}+\varphi) \leq 1$, the convex/concave quadratic surrogate function of $\hat{f}(\mathbf{q} \vert \mathbf{k},\varphi)$ w.r.t. $\mathbf{q}$ can be constructed at point $\mathbf{q}_{0}$ to globally majorize/minorize it, i.e.,
\begin{subequations}\label{eq_surrogate_cos}
	\begin{align}
		&\begin{aligned}
			&\hat{f}(\mathbf{q} \vert \mathbf{k},\varphi) \leq  \cos(\mathbf{k}^{\mathrm{T}}\mathbf{q}_{0}+\varphi) - \sin(\mathbf{k}^{\mathrm{T}}\mathbf{q}_{0}+\varphi)\mathbf{k}^{\mathrm{T}}(\mathbf{q} - \mathbf{q}_{0})\\
			& +\frac{1}{2} \left(\mathbf{k}^{\mathrm{T}}(\mathbf{q} - \mathbf{q}_{0})\right)^2 \triangleq \hat{f}_{\mathrm{ub}} \left( \mathbf{q} \vert \mathbf{q}_{0}, \mathbf{k},\varphi \right), 
		\end{aligned}\label{eq_surrogate_cos_up}\\
		&\begin{aligned}
			&\hat{f}(\mathbf{q} \vert \mathbf{k},\mu) \geq  \cos(\mathbf{k}^{\mathrm{T}}\mathbf{q}_{0}+\varphi) - \sin(\mathbf{k}^{\mathrm{T}}\mathbf{q}_{0}+\varphi)\mathbf{k}^{\mathrm{T}}(\mathbf{q} - \mathbf{q}_{0})\\
			& -\frac{1}{2} \left(\mathbf{k}^{\mathrm{T}}(\mathbf{q} - \mathbf{q}_{0})\right)^2 \triangleq \hat{f}_{\mathrm{lb}} \left( \mathbf{q} \vert \mathbf{q}_{0}, \mathbf{k},\varphi \right). 
		\end{aligned}\label{eq_surrogate_cos_low}
	\end{align}
\end{subequations}

\begin{figure*}[!t]
	\centering 
	\begin{equation}\label{eq_interfer_ave_bound}
		\begin{aligned}
			&I(\mathbf{q}, \mathbf{w}^{(i-1)}) = \sum \limits_{m=1}^{M} I_{m}(\mathbf{q}[m], \mathbf{w}^{(i-1)}[m]) 
			= \sum \limits_{m=1}^{M}\sum \limits_{l=1}^{L_m} g_{m, l}^{\mathrm{int}}\left| \sum \limits_{n=1}^{N} w_{n}^{(i-1)}[m] e^{-j {\mathbf{k}_{m, l}^{\mathrm{int}}}^{\mathrm{T}} \mathbf{q}_{n}[m]} \right|^2\\
			= &\frac{1}{N} \sum \limits_{m=1}^{M}\sum \limits_{l=1}^{L_m} g_{m, l}^{\mathrm{int}} \sum \limits_{n=1}^{N} \sum \limits_{\hat{n}=1}^{N} \cos \left\{
			{\mathbf{k}_{m, l}^{\mathrm{int}}}^{\mathrm{T}} \left(\mathbf{q}_{n}[m] - \mathbf{q}_{\hat{n}}[m]\right) + \varphi_{\hat{n}}^{(i-1)}[m] - \varphi_{n}^{(i-1)}[m] \right\}\\
			\leq & \frac{1}{N} \sum \limits_{m=1}^{M}\sum \limits_{l=1}^{L_m} g_{m, l}^{\mathrm{int}} \sum \limits_{n=1}^{N} \sum \limits_{\hat{n}=1}^{N} \hat{f}_{\mathrm{ub}}\left(\mathbf{q}_{n}[m] - \mathbf{q}_{\hat{n}}[m] ~{\Big\vert}~ \mathbf{q}_{n}^{(i-1)}[m] - \mathbf{q}_{\hat{n}}^{(i-1)}[m], \mathbf{k}_{m, l}^{\mathrm{int}}, \varphi_{\hat{n}}^{(i-1)}[m] - \varphi_{n}^{(i-1)}[m] \right)\\
			\triangleq & \hat{I}(\mathbf{q} ~{\big\vert}~ \mathbf{q}^{(i-1)}, \mathbf{w}^{(i-1)}).
		\end{aligned}
	\end{equation}
	\vspace*{8pt}
	\hrulefill
	\begin{equation}\label{eq_beamgain_ave_bound}
		\begin{aligned}
			&G_{m}(\mathbf{q}[m], \mathbf{w}^{(i-1)}[m]) = \sum \limits_{l=1}^{L_0} g_{m, l}^{\mathrm{cov}}\left| \sum \limits_{n=1}^{N} w_{n}^{(i-1)}[m] e^{-j {\mathbf{k}_{m, l}^{\mathrm{cov}}}^{\mathrm{T}} \mathbf{q}_{n}[m]} \right|^2\\
			= &\frac{1}{N} \sum \limits_{l=1}^{L_0} g_{m, l}^{\mathrm{cov}} \sum \limits_{n=1}^{N} \sum \limits_{\hat{n}=1}^{N} \cos \left\{
			{\mathbf{k}_{m, l}^{\mathrm{cov}}}^{\mathrm{T}} \left(\mathbf{q}_{n}[m] - \mathbf{q}_{\hat{n}}[m]\right) + \varphi_{\hat{n}}^{(i-1)}[m] - \varphi_{n}^{(i-1)}[m] \right\}\\
			\geq & \frac{1}{N} \sum \limits_{l=1}^{L_0} g_{m, l}^{\mathrm{cov}} \sum \limits_{n=1}^{N} \sum \limits_{\hat{n}=1}^{N} \hat{f}_{\mathrm{lb}}\left(\mathbf{q}_{n}[m] - \mathbf{q}_{\hat{n}}[m] ~{\Big\vert}~ \mathbf{q}_{n}^{(i-1)}[m] - \mathbf{q}_{\hat{n}}^{(i-1)}[m], \mathbf{k}_{m, l}^{\mathrm{cov}}, \varphi_{\hat{n}}^{(i-1)}[m] - \varphi_{n}^{(i-1)}[m] \right)\\
			\triangleq & \hat{G}_{m}(\mathbf{q}[m] ~{\big\vert}~ \mathbf{q}^{(i-1)}[m], \mathbf{w}^{(i-1)}[m]).
		\end{aligned}
	\end{equation}
	\vspace*{8pt}
	\hrulefill
\end{figure*}

Denote $\mathbf{q}^{(i-1)}$ as the solution for APV obtained in the $(i-1)$-th iteration and $\mathbf{w}^{(i-1)}=e^{j \boldsymbol{\varphi}^{(i-1)}}/\sqrt{N}$ with $\boldsymbol{\varphi}^{(i-1)}$ being the phase vector of $\mathbf{w}^{(i-1)}$. By utilizing \eqref{eq_surrogate_cos_up}, the objective function in \eqref{eq_problem_APV_a} can be globally upper-bounded by its convex quadratic surrogate function $\hat{I}(\mathbf{q} ~{\vert} \mathbf{q}^{(i-1)}, \mathbf{w}^{(i-1)})$ w.r.t. $\mathbf{q}$ shown in \eqref{eq_interfer_ave_bound} at the top of the next page. Similarly, by utilizing \eqref{eq_surrogate_cos_low}, $G_{m}(\mathbf{q}[m], \mathbf{w}^{(i-1)}[m])$ in \eqref{eq_problem_APV_b} can be globally lower-bounded by its convex quadratic surrogate function $\hat{G}_{m}(\mathbf{q}[m] ~{\vert} \mathbf{q}^{(i-1)}[m], \mathbf{w}^{(i-1)}[m])$ w.r.t. $\mathbf{q}[m]$ shown in \eqref{eq_beamgain_ave_bound} at the top of the next page. 

Next, we handle the non-convex constraint \eqref{eq_problem_APV_d}. Since $\left\|\mathbf{q}_{n}[m]-\mathbf{q}_{\hat{n}}[m]\right\|_{2}$ is a convex function w.r.t. $(\mathbf{q}_{n}[m]-\mathbf{q}_{\hat{n}}[m])$, it can be globally lower-bounded by the first-order Taylor expansion at $(\mathbf{q}_{n}^{(i-1)}[m]-\mathbf{q}_{\hat{n}}^{(i-1)}[m])$ as follows:
	\begin{equation}\label{eq_interMA_dis}
	\begin{aligned}
		&\left\|\mathbf{q}_{n}[m]-\mathbf{q}_{\hat{n}}[m]\right\|_{2} \geq \left\|\mathbf{q}_{n}^{(i-1)}[m]-\mathbf{q}_{\hat{n}}^{(i-1)}[m]\right\|_{2}\\
		 &+\frac{(\mathbf{q}_{n}^{(i-1)}[m]-\mathbf{q}_{\hat{n}}^{(i-1)}[m])^{\mathrm{T}}}{\left\|\mathbf{q}_{n}^{(i-1)}[m]-\mathbf{q}_{\hat{n}}^{(i-1)}[m]\right\|_{2}} \cdot\\ &~~~~~~~~~\left\{(\mathbf{q}_{n}[m]-\mathbf{q}_{\hat{n}}[m])-(\mathbf{q}_{n}^{(i-1)}[m]-\mathbf{q}_{\hat{n}}^{(i-1)}[m])\right\}\\
		 &=\frac{(\mathbf{q}_{n}^{(i-1)}[m]-\mathbf{q}_{\hat{n}}^{(i-1)}[m])^{\mathrm{T}}(\mathbf{q}_{n}[m]-\mathbf{q}_{\hat{n}}[m])}{\left\|\mathbf{q}_{n}^{(i-1)}[m]-\mathbf{q}_{\hat{n}}^{(i-1)}[m]\right\|_{2}}\\
		 &\triangleq \hat{d}(\mathbf{q}_{n}[m],\mathbf{q}_{\hat{n}}[m]).
	\end{aligned}
\end{equation}

As such, problem \eqref{eq_problem_APV} can be relaxed as 
\begin{subequations}\label{eq_problem_APV2}
	\begin{align}
		\mathop{\min}\limits_{\mathbf{q}}~
		&\hat{I}(\mathbf{q} ~{\big\vert} \mathbf{q}^{(i-1)}, \mathbf{w}^{(i-1)}) \label{eq_problem_APV2_a}\\
		\mathrm{s.t.}~~ & \hat{G}_{m}(\mathbf{q}[m] ~{\big\vert} \mathbf{q}^{(i-1)}[m], \mathbf{w}^{(i-1)}[m]) \geq \eta,~ 1 \leq m \leq M, \label{eq_problem_APV2_b}\\
		&\hat{d}(\mathbf{q}_{n}[m],\mathbf{q}_{\hat{n}}[m]) \geq d_{\min},~1 \leq m \leq M, n \neq \hat{n}, \label{eq_problem_APV2_c}\\
		&\eqref{eq_problem_APV_c},~\eqref{eq_problem_APV_e},  \label{eq_problem_APV2_d}
	\end{align}
\end{subequations}
which is a convex optimization problem with a quadratic objective function and quadratic constraints. Thus, it can be solved by using the existing quadratic optimization tools such as CVX \cite{boyd2004convex}. It is wroth noting that the dimension of $\mathbf{q}$ is $2MN$, which is large in general. To reduce the computational complexity, we adopt the block coordinate descent (BCD) technique to successively optimize the subvectors of $\mathbf{q}$. Specifically, the $M$ time slots are divided into multiple groups with each group including $M_{0} (\leq M)$ time slots. The APVs for the $k$-th group of time slots, i.e., $\mathbf{q}[(k-1)M_{0}+1], \mathbf{q}[(k-1)M_{0}+2], \dots \mathbf{q}[kM_{0}]$, are successively optimized for $k=1,2,\cdots, M/M_{0}$. As such, the dimension of optimization variables is reduced to $2M_{0}N$ for each iteration.

\subsection{Optimization of AWV}
Note that for any given $\mathbf{q}^{(i)}$ obtained in the $i$-th iteration, the optimization of AWV is independent over different time slots. As such, the AWV at the $m$-th time slot, $1 \leq m \leq M$, can be optimized by solving the following subproblem,
\begin{subequations}\label{eq_problem_AWV}
	\begin{align}
		\mathop{\min}\limits_{\mathbf{w}[m]}~
		&I_{m}(\mathbf{q}^{(i)}[m], \mathbf{w}[m]) \label{eq_problem_AWV_a}\\
		\mathrm{s.t.}~~ & G_{m}(\mathbf{q}^{(i)}[m], \mathbf{w}[m]) \geq \eta, \label{eq_problem_AWV_b}\\
		&\left|w_{n}[m]\right|=\frac{1}{\sqrt{N}},~1 \leq n \leq N.  \label{eq_problem_AWV_c}
	\end{align}
\end{subequations}
To handle the constant-modulus constraint in \eqref{eq_problem_AWV_c}, we define $\mathbf{w}[m]=e^{j \boldsymbol{\varphi}[m]}/\sqrt{N}$ and optimize the phase vector $\boldsymbol{\varphi}[m] \in \mathbb{R}^{N \times 1}$ instead. Since $I_{m}(\mathbf{q}^{(i)}[m], e^{j \boldsymbol{\varphi}[m]}/\sqrt{N})$ and $G_{m}(\mathbf{q}^{(i)}[m], e^{j \boldsymbol{\varphi}[m]}/\sqrt{N})$ are both non-convex/non-concave functions w.r.t. $\boldsymbol{\varphi}[m]$, we construct surrogate functions for them at local point $\boldsymbol{\varphi}^{(i-1)}[m]=\angle \mathbf{w}^{(i-1)}[m]$. 

To this end, for any real constant vectors $\mathbf{k}$ and $\mathbf{q}$, we derive the second-order Taylor expansion of function $\bar{f}(\varphi \vert \mathbf{k}^{\mathrm{T}}\mathbf{q}) = \cos(\varphi + \mathbf{k}^{\mathrm{T}}\mathbf{q})$ at any given point $\varphi_{0}$ as 
\begin{equation}\label{eq_Taylor_cos2}
	\begin{aligned}	
		&\cos(\varphi_{0} + \mathbf{k}^{\mathrm{T}}\mathbf{q}) - \sin(\varphi_{0} + \mathbf{k}^{\mathrm{T}}\mathbf{q})(\varphi - \varphi_{0})\\
		& ~~~~~~~~~~~~~~~~~~~-\frac{1}{2} \cos(\varphi_{0} + \mathbf{k}^{\mathrm{T}}\mathbf{q}) (\varphi - \varphi_{0})^{2}.
	\end{aligned}
\end{equation}
Given the fact of $-1 \leq \cos(\varphi_{0} + \mathbf{k}^{\mathrm{T}}\mathbf{q}) \leq 1$, the convex/concave quadratic surrogate function of $\bar{f}(\varphi \vert \mathbf{k}, \mathbf{q})$ w.r.t. $\varphi$ can be constructed at point $\varphi$ to globally majorize/minorize it, i.e.,
\begin{subequations}\label{eq_surrogate_cos2}
	\begin{align}
		&\begin{aligned}
			\bar{f}(\varphi \vert \mathbf{k}^{\mathrm{T}}\mathbf{q}) \leq  &\cos(\varphi_{0} + \mathbf{k}^{\mathrm{T}}\mathbf{q}) - \sin(\varphi_{0} + \mathbf{k}^{\mathrm{T}}\mathbf{q})(\varphi - \varphi_{0})\\
			& +\frac{1}{2} (\varphi - \varphi_{0})^{2} \triangleq \bar{f}_{\mathrm{ub}} \left(\varphi \vert \varphi_{0}, \mathbf{k}^{\mathrm{T}}\mathbf{q} \right), 
		\end{aligned}\label{eq_surrogate_cos_up2}\\
		&\begin{aligned}
			\bar{f}(\varphi \vert \mathbf{k}^{\mathrm{T}}\mathbf{q}) \geq  &\cos(\varphi_{0} + \mathbf{k}^{\mathrm{T}}\mathbf{q}) - \sin(\varphi_{0} + \mathbf{k}^{\mathrm{T}}\mathbf{q})(\varphi - \varphi_{0})\\
			& -\frac{1}{2} (\varphi - \varphi_{0})^{2} \triangleq \bar{f}_{\mathrm{lb}} \left(\varphi \vert \varphi_{0}, \mathbf{k}^{\mathrm{T}}\mathbf{q} \right).
		\end{aligned}\label{eq_surrogate_cos_low2}
	\end{align}
\end{subequations}

By utilizing \eqref{eq_surrogate_cos_up2}, the average interference leakage power in \eqref{eq_problem_AWV_a} can be globally upper-bounded by its convex quadratic surrogate function $\bar{I}_{m}(\boldsymbol{\varphi}[m] ~{\vert} \mathbf{q}^{(i)}[m], \boldsymbol{\varphi}^{(i-1)}[m])$ w.r.t. $\boldsymbol{\varphi}[m]$ shown in \eqref{eq_interfer_ave_bound2} at the top of the next page. Similarly, by utilizing \eqref{eq_surrogate_cos_low2}, the average beamforming gain in \eqref{eq_problem_AWV_b} can be globally lower-bounded by its convex quadratic surrogate function $\bar{G}_{m}(\boldsymbol{\varphi}[m] ~{\vert} \mathbf{q}^{(i-1)}[m], \boldsymbol{\varphi}^{(i-1)}[m])$ w.r.t. $\boldsymbol{\varphi}[m]$ shown in \eqref{eq_beamgain_ave_bound2} at the top of the next page.
\begin{figure*}[!t]
	\centering 
	\begin{equation}\label{eq_interfer_ave_bound2}
		\begin{aligned}
			&I_{m}(\mathbf{q}^{(i)}[m], e^{j \boldsymbol{\varphi}[m]}/\sqrt{N}) = 
			\frac{1}{N} \sum \limits_{l=1}^{L_m} g_{m, l}^{\mathrm{int}}\left| \sum \limits_{n=1}^{N}  e^{-j \left\{ \varphi_{n}[m]-{\mathbf{k}_{m, l}^{\mathrm{int}}}^{\mathrm{T}} \mathbf{q}_{n}^{(i)}[m]\right\}} \right|^2\\
			= &\frac{1}{N} \sum \limits_{l=1}^{L_m} g_{m, l}^{\mathrm{int}} \sum \limits_{n=1}^{N} \sum \limits_{\hat{n}=1}^{N} \cos \left\{
			\varphi_{n}[m] - \varphi_{\hat{n}}[m] + {\mathbf{k}_{m, l}^{\mathrm{int}}}^{\mathrm{T}} \left(\mathbf{q}_{\hat{n}}^{(i)}[m] - \mathbf{q}^{(i)}_{n}[m]\right)  \right\}\\
			\leq & \frac{1}{N} \sum \limits_{l=1}^{L_m} g_{m, l}^{\mathrm{int}} \sum \limits_{n=1}^{N} \sum \limits_{\hat{n}=1}^{N} \bar{f}_{\mathrm{ub}}\left(\varphi_{n}[m] - \varphi_{\hat{n}}[m] ~{\Big\vert}~ \varphi_{n}^{(i-1)}[m] - \varphi_{\hat{n}}^{(i-1)}[m], {\mathbf{k}_{m, l}^{\mathrm{int}}}^{\mathrm{T}} \left(\mathbf{q}_{\hat{n}}^{(i)}[m] - \mathbf{q}^{(i)}_{n}[m]\right)\right)\\
			\triangleq & \bar{I}_{m}(\boldsymbol{\varphi}[m] ~{\big\vert}~ \mathbf{q}^{(i)}[m], \boldsymbol{\varphi}^{(i-1)}[m]).
		\end{aligned}
	\end{equation}
	\vspace*{8pt}
	\hrulefill
	\begin{equation}\label{eq_beamgain_ave_bound2}
		\begin{aligned}
			&G_{m}(\mathbf{q}^{(i)}[m], e^{j \boldsymbol{\varphi}[m]}/\sqrt{N}) = 
			\frac{1}{N} \sum \limits_{l=1}^{L_m} g_{m, l}^{\mathrm{cov}}\left| \sum \limits_{n=1}^{N}  e^{-j \left\{ \varphi_{n}[m]-{\mathbf{k}_{m, l}^{\mathrm{cov}}}^{\mathrm{T}} \mathbf{q}_{n}^{(i)}[m]\right\}} \right|^2\\
			= &\frac{1}{N} \sum \limits_{l=1}^{L_m} g_{m, l}^{\mathrm{cov}} \sum \limits_{n=1}^{N} \sum \limits_{\hat{n}=1}^{N} \cos \left\{
			\varphi_{n}[m] - \varphi_{\hat{n}}[m] + {\mathbf{k}_{m, l}^{\mathrm{cov}}}^{\mathrm{T}} \left(\mathbf{q}_{\hat{n}}^{(i)}[m] - \mathbf{q}^{(i)}_{n}[m]\right)  \right\}\\
			\geq & \frac{1}{N} \sum \limits_{l=1}^{L_m} g_{m, l}^{\mathrm{cov}} \sum \limits_{n=1}^{N} \sum \limits_{\hat{n}=1}^{N} \bar{f}_{\mathrm{lb}}\left(\varphi_{n}[m] - \varphi_{\hat{n}}[m] ~{\Big\vert}~ \varphi_{n}^{(i-1)}[m] - \varphi_{\hat{n}}^{(i-1)}[m], {\mathbf{k}_{m, l}^{\mathrm{cov}}}^{\mathrm{T}} \left(\mathbf{q}_{\hat{n}}^{(i)}[m] - \mathbf{q}^{(i)}_{n}[m]\right)\right)\\
			\triangleq & \bar{G}_{m}(\boldsymbol{\varphi}[m] ~{\big\vert}~ \mathbf{q}^{(i)}[m], \boldsymbol{\varphi}^{(i-1)}[m]).
		\end{aligned}
	\end{equation}
	\vspace*{8pt}
	\hrulefill
\end{figure*}

As such, problem \eqref{eq_problem_AWV} can be relaxed as 
\begin{subequations}\label{eq_problem_AWV2}
	\begin{align}
		\mathop{\min}\limits_{\boldsymbol{\varphi}[m]}~
		&\bar{I}_{m}(\boldsymbol{\varphi}[m] ~{\big\vert}~ \mathbf{q}^{(i)}[m], \boldsymbol{\varphi}^{(i-1)}[m]) \label{eq_problem_AWV2_a}\\
		\mathrm{s.t.}~~ & \bar{G}_{m}(\boldsymbol{\varphi}[m] ~{\big\vert}~ \mathbf{q}^{(i)}[m], \boldsymbol{\varphi}^{(i-1)}[m]) \geq \eta, \label{eq_problem_AWV2_b}
	\end{align}
\end{subequations}
which is a convex quadratic optimization problem and can be solved by CVX \cite{boyd2004convex}.

\subsection{Overall Algorithm}
\begin{algorithm}[t]
	\caption{Solution for solving problem \eqref{eq_problem_dis}.}
	\label{alg_proposed}
	\begin{algorithmic}[1]
		\REQUIRE ~$\bar{\mathcal{A}}_{\mathrm{e}}$, $\{\bar{\mathcal{A}}_{\mathrm{i}}^{m}\}$, $T$, $M$, $N$, $\eta$, $\mathcal{C}$, $d_{\min}$, $v_{\max}$,\\
		~~~~~~~$M_{0}$, $I_{\max}$, $\epsilon$.
		\ENSURE ~$\mathbf{q}^{\star}$, $\mathbf{w}^{\star}$. \\
		\STATE Initialize $\mathbf{q}^{(0)}$ and $\mathbf{w}^{(0)}$ according to \eqref{eq_AWV_ini}.
		\STATE Calculate $\boldsymbol{\varphi}^{(0)} = \angle \mathbf{w}^{(0)}$.
		\FOR   {$i=1:1:I_{\max}$}
		\FOR   {$k=1:1:M/M_{0}$}
		\STATE Update $\{\mathbf{q}^{(i)}[(k-1)M_{0}+1],\dots \mathbf{q}^{(i)}[kM_{0}]\}$ by solving problem \eqref{eq_problem_APV2}.
		\ENDFOR 
		\FOR   {$m=1:1:M$}
		\STATE Update $\boldsymbol{\varphi}^{(i)}[m]$ by solving problem \eqref{eq_problem_AWV2}.
		\STATE Update $\mathbf{w}^{(i)}[m] = e^{j \boldsymbol{\varphi}^{(i)}[m]}/\sqrt{N}$.
		\ENDFOR 
		\IF {$|I(\mathbf{q}^{(i-1)}, \mathbf{w}^{(i-1)}) - I(\mathbf{q}^{(i)}, \mathbf{w}^{(i)})| \leq \epsilon$ }
		\STATE Break.
		\ENDIF
		\ENDFOR
		\STATE Set APV and AWV as $\mathbf{q}^{\star}=\mathbf{q}^{(i)}$ and $\mathbf{w}^{\star}=\mathbf{w}^{(i)}$.
		\RETURN $\mathbf{q}^{\star}$, $\mathbf{w}^{\star}$.
	\end{algorithmic}
\end{algorithm} 

The overall algorithm for alternately optimizing the APV, $\mathbf{q}$, and the AWV, $\mathbf{w}$, is summarized in Algorithm \ref{alg_proposed}. The initialized APV, $\mathbf{q}^{(0)}$, is given by adopting the UPA geometry with half-wavelength antenna spacing for all time slots. The AWV is initialized by using the normalized steering vector pointing to the coverage center $(\Theta_{\kappa}, \Phi_{\kappa})$ for each time slot, i.e., 
\begin{equation}\label{eq_AWV_ini}
	\mathbf{w}^{(0)}[m] = \frac{1}{\sqrt{N}}\mathbf{a}(\mathbf{k}_{m, \kappa}^{\mathrm{cov}}, \mathbf{q}^{(0)}[m]), 1 \leq m \leq M.
\end{equation}
The APVs, $\{\mathbf{q}^{(i)}[(k-1)M_{0}+1],\dots \mathbf{q}^{(i)}[kM_{0}]\}$, at different time slots are successively updated in lines 4-6 to decrease the computational complexity. The AWVs, $\mathbf{w}^{(i)}[m]$, at different time slots are separately updated in lines 7-10. The algorithm terminates if the decrease of the objective function is below a predefined threshold $\epsilon$ or the iteration index exceeds its maximum value given by $I_{\max}$. 

In addition, the following inequality equalities/inequalities hold over the iterations:
\begin{equation}\label{eq_convergence}
	\begin{aligned}
		&I(\mathbf{q}^{(i-1)}, \mathbf{w}^{(i-1)}) \\
		\overset{\text{(a)}}{=}&\hat{I}(\mathbf{q}^{(i-1)} ~{\big\vert} \mathbf{q}^{(i-1)}, \mathbf{w}^{(i-1)}) \\
		\overset{\text{(b)}}{\geq} &\hat{I}(\mathbf{q}^{(i)} ~{\big\vert} \mathbf{q}^{(i-1)}, \mathbf{w}^{(i-1)}) \\
		\overset{\text{(c)}}{\geq} &I(\mathbf{q}^{(i)}, \mathbf{w}^{(i-1)}) = I(\mathbf{q}^{(i)}, e^{j \boldsymbol{\varphi}^{(i-1)}}/\sqrt{N})\\
		\overset{\text{(d)}}{=} &\sum \limits_{m=1}^{M} \bar{I}_{m}(\boldsymbol{\varphi}^{(i-1)}[m] ~{\big\vert}~ \mathbf{q}^{(i)}[m], \boldsymbol{\varphi}^{(i-1)}[m])\\
		\overset{\text{(e)}}{\geq} &\sum \limits_{m=1}^{M} \bar{I}_{m}(\boldsymbol{\varphi}^{(i)}[m] ~{\big\vert}~ \mathbf{q}^{(i)}[m], \boldsymbol{\varphi}^{(i-1)}[m])\\
		\overset{\text{(f)}}{\geq} &I(\mathbf{q}^{(i)}, \mathbf{w}^{(i)}),
	\end{aligned}
\end{equation}
where equality (a) holds because of $\hat{f}(\mathbf{q} \vert \mathbf{k},\varphi) = \hat{f}_{\mathrm{ub}} \left( \mathbf{q} \vert \mathbf{q}_{0}, \mathbf{k},\varphi \right)$ at $\mathbf{q} = \mathbf{q}_{0}$ according to \eqref{eq_surrogate_cos_up}; inequality (b) holds because $\mathbf{q}^{(i)}$ is the optimal solution yielding the minimum objective function for problem \eqref{eq_problem_APV2}; inequality (c) holds according to \eqref{eq_interfer_ave_bound}; equality (d) holds because $\bar{f}(\varphi \vert \mathbf{k}^{\mathrm{T}}\mathbf{q}) = \bar{f}_{\mathrm{ub}} \left(\varphi \vert \varphi_{0}, \mathbf{k}^{\mathrm{T}}\mathbf{q} \right)$ at $\varphi = \varphi_{0}$ according to \eqref{eq_surrogate_cos_up2}; inequality (e) holds because $\boldsymbol{\varphi}^{(i)}[m]$ is the optimal solution yielding the minimum objective function for problem \eqref{eq_problem_AWV2}; and inequality (f) holds according to \eqref{eq_interfer_ave_bound2}. Since the objective function $I(\mathbf{q}, \mathbf{w})$ is positive and it is non-increasing with the iteration index, the convergence of Algorithm \ref{alg_proposed} is thus guaranteed.

The computational complexity of Algorithm \ref{alg_proposed} is analyzed as follows. The computational complexity of solving the APVs at $M_{0}$ time slots in problem \eqref{eq_problem_APV2} is $\mathcal{O}(M_{0}^{3.5}N^{3.5})$. Thus, the complexity for executing lines 4-6 is $\mathcal{O}(MM_{0}^{2.5}N^{3.5})$. The computational complexity of solving the AWV at each time slot in problem \eqref{eq_problem_AWV2} is $\mathcal{O}(N^{3.5})$, which entails a complexity of $\mathcal{O}(MN^{3.5})$ for executing lines 7-10. As a result, the total computational complexity of Algorithm \ref{alg_proposed} is $\mathcal{O}(I_{\max}MM_{0}^{2.5}N^{3.5})$.

\subsection{Low-Complexity Scheme}
The proposed solution in Algorithm \ref{alg_proposed} involves the optimization of APVs at all time slots. As a result, each MA element may need to change its position multiple times within the entire interval $[0, T)$, which incurs a high movement overhead and energy consumption. To resolve this issue, we propose in this subsection a low-complexity MA (LC-MA) movement scheme, where an optimized common MA array geometry is used over all the $M$ time slots. As such, the MA array only needs to configure its geometry at the beginning of each time interval based on the coverage requirement, which can significantly reduce the antenna movement overhead. Defining $\mathbf{q}[1]=\dots=\mathbf{q}[M]=\tilde{\mathbf{q}} \in \mathbb{R}^{2N \times 1}$ as the common APV, problem \eqref{eq_problem_dis} is thus recast to
\begin{subequations}\label{eq_problem_dis_LC}
	\begin{align}
		\mathop{\min}\limits_{\tilde{\mathbf{q}}, \mathbf{w}}~
		&I(\mathbf{1}_{M \times 1} \otimes \tilde{\mathbf{q}}, \mathbf{w}) \label{eq_problem_dis_LC_a}\\
		\mathrm{s.t.}~~ & G_{m}(\tilde{\mathbf{q}}, \mathbf{w}[m]) \geq \eta,~ 1 \leq m \leq M, \label{eq_problem_dis_LC_b}\\
		&\tilde{\mathbf{q}}_n \in \mathcal{C}, ~ 1 \leq n \leq N,  \label{eq_problem_dis_LC_c}\\
		&\left\|\tilde{\mathbf{q}}_n-\tilde{\mathbf{q}}_{\hat{n}}\right\|_2 \geq d_{\min },~ n \neq \hat{n}, \label{eq_problem_dis_LC_d}\\
		&\left|w_{n}[m]\right|=\frac{1}{\sqrt{N}},~1 \leq m \leq M, 1 \leq n \leq N. \label{eq_problem_dis_LC_e}
	\end{align}
\end{subequations}

Problem \eqref{eq_problem_dis_LC} can be solved by alternately optimizing $\tilde{\mathbf{q}}$ and $\mathbf{w}$ similarly to that in Algorithm \ref{alg_proposed}, whereas the optimization of APV can be simplified as
\begin{subequations}\label{eq_problem_APV_LC}
	\begin{align}
		\mathop{\min}\limits_{\tilde{\mathbf{q}}}~
		&I(\mathbf{1}_{M \times 1} \otimes \tilde{\mathbf{q}}, \mathbf{w}^{(i-1)}) \label{eq_problem_APV_LC_a}\\
		\mathrm{s.t.}~~ & G_{m}(\tilde{\mathbf{q}}, \mathbf{w}^{(i-1)}[m]) \geq \eta,~ 1 \leq m \leq M, \label{eq_problem_APV_LC_b}\\
		&\tilde{\mathbf{q}}_n \in \mathcal{C}, ~ 1 \leq n \leq N,  \label{eq_problem_APV_LC_c}\\
		&\left\|\tilde{\mathbf{q}}_n-\tilde{\mathbf{q}}_{\hat{n}}\right\|_2 \geq d_{\min },~n \neq \hat{n}, \label{eq_problem_APV_LC_d}
	\end{align}
\end{subequations}
which can be solved by using the SCA technique similar to that in Section III-B. The corresponding computational complexity for optimizing the common APV is reduced to $\mathcal{O}((M+N)^{3.5})$. Thus, the total computational complexity of the LC-MA scheme is given by $\mathcal{O}((M+N)^{3.5}+I_{\max}MN^{3.5})$, which is lower than that of Algorithm \ref{alg_proposed}.

\section{Simulation Results}
In this section, we present simulation results to evaluate the performance of the proposed MA array-aided dynamic beam coverage strategy for LEO satellite communications and compare it to other benchmark schemes. The simulation setup and benchmark schemes are first introduced and then the numerical results are presented.

\subsection{Simulation Setup and Benchmark Schemes}
In the simulation, we consider a typical Walker Delta LEO constellation with orbit altitude $H_{\mathrm{s}}=1500$ kilometers (km) and inclination angle $65^{\circ}$. The number of satellites in each orbital plane is set to $K_{\mathrm{s}}=24$. Thus, the length of the considered time interval can be obtained as $T =T_{\mathrm{s}}/K_{\mathrm{s}} =289.56$ seconds (s). The initial angle between the directions of the ascending node and the satellite at time $t=0$ is set to $\alpha_{0}=-7.5^{\circ}$, which indicates $\alpha(T)=7.5^{\circ}$. The coverage area on the earth surface is defined as a circle centered at $(\Theta, \Phi)=(0,0)$, where the geocentric angle between the coverage center and border is $3^{\circ}$. Note that for this setup, the trajectory of the satellite and the coverage area are both centrosymmetric w.r.t. the coverage center. The carrier frequency is set to 14 GHz and the path loss exponent is set to $\gamma=2.8$. The number of discrete time slots is $M=50$. The number of quantized coordinates $(\Theta, \Phi)$ is set to $L_{\mathrm{e}} \times L_{\mathrm{a}} = 100 \times 200$. The number of MAs is set to $N=16$. The threshold for the minimum beamforming gain over the coverage area is $\eta = 0.5N$. The antenna moving region $\mathcal{C}$ is set as a square area of size $3\lambda \times 3\lambda$. The minimum inter-MA distance is set to $d_{\min}=\lambda/2$ and the maximum moving speed of each MA is $v_{\max}=0.01$ meters per second (m/s). The maximum number of iterations and the termination threshold in Algorithm \ref{alg_proposed} are set to $I_{\max}=1000$ and $\epsilon = 10^{-4}$, respectively. The proposed and benchmark schemes are defined as follows.

\begin{itemize}
	\item UPA, steering: An UPA of square size with $4 \times 4$ antennas and half-wavelength inter-antenna spacing is adopted at the satellite, where the normalized steering vector pointing to the coverage center in \eqref{eq_AWV_ini} is used for beamforming at each time slot. 
	\item UPA, optimized: The same UPA as above is adopted, where the AWV at each time slot is optimized by the method given in Section III-C.
	\item MA: The proposed MA beamforming solution based on the AO over the APV and AWV, as given in Section III-D.
	\item 6DMA: In addition to the APV and AWV optimization, the array orientation/rotation is also optimized at each time slot by the AO. This scheme can be considered as a special realization of 6DMA \cite{shao20246DMA,shao2024discrete} with all antennas equipped on a single movable surface for orientation/rotation adjustment.
	\item LC-MA: The proposed LC-MA beamforming solution based on the AO over the common APV for all time slots and the AWVs for different time slots, as given in Section III-E.
	\item LC-6DMA: Based on the LC-MA, the array orientations/rotations for different time slots are also optimized jointly with the common APV and the AWVs via the AO.
\end{itemize}

For all the above schemes, we assume that each antenna element adopts an half-space isotropic radiation pattern pointing downwards given by
\begin{equation}
	G_{\mathrm{AE}}\left(\mathbf{k}(\Theta, \Phi, t)\right) = \left\{\begin{aligned}
		& 1, ~ \text{if} \left[\mathbf{k}(\Theta, \Phi, t)\right]_{3} > 0,\\
		& 0, ~ \text{otherwise}.
		\end{aligned}\right.
\end{equation}

\begin{figure}[t]
	\centering
	\includegraphics[width=\figwidth cm]{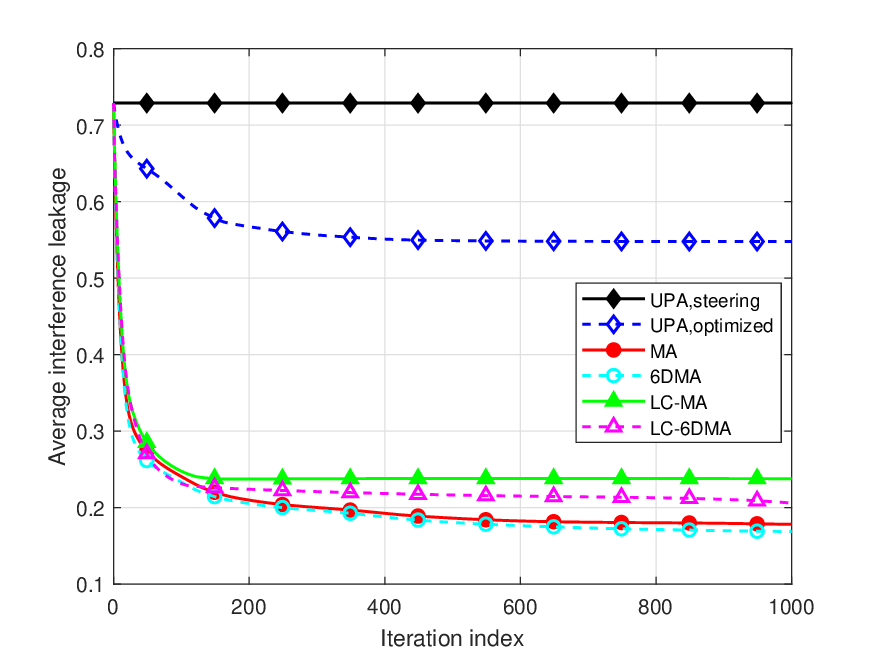}
	\caption{The convergence of the proposed algorithms and benchmark schemes.}
	\label{Fig_ite}
\end{figure}
\begin{figure}[t]
	\centering
	\includegraphics[width=\figwidth cm]{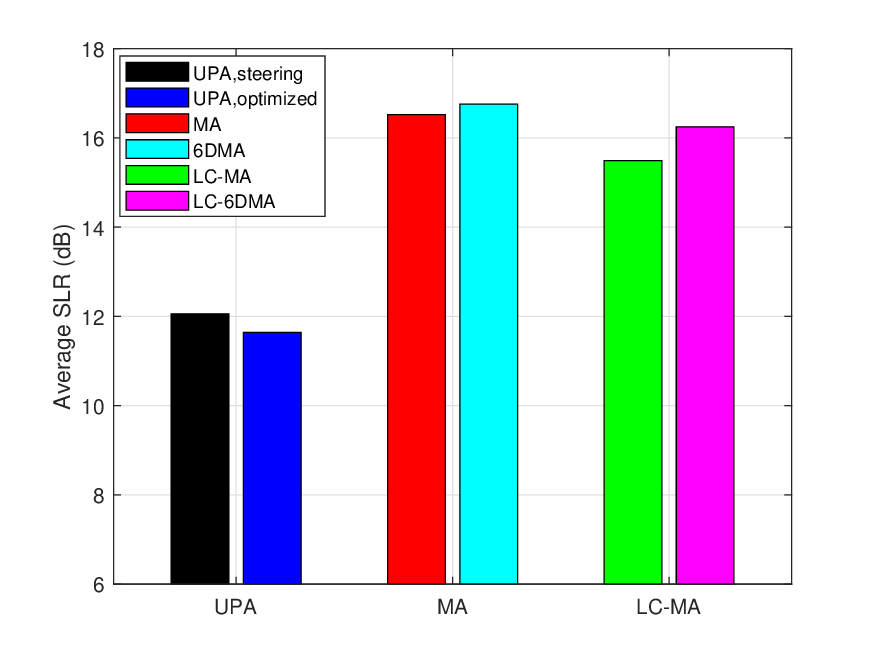}
	\caption{Comparison of the average SLR between the proposed and benchmark schemes.}
	\label{Fig_SLR}
\end{figure}

\subsection{Numerical Results}
\begin{figure*}[t]
	\centering
	\subfigure[UPA, steering, $m=1$]{\includegraphics[width=5 cm]{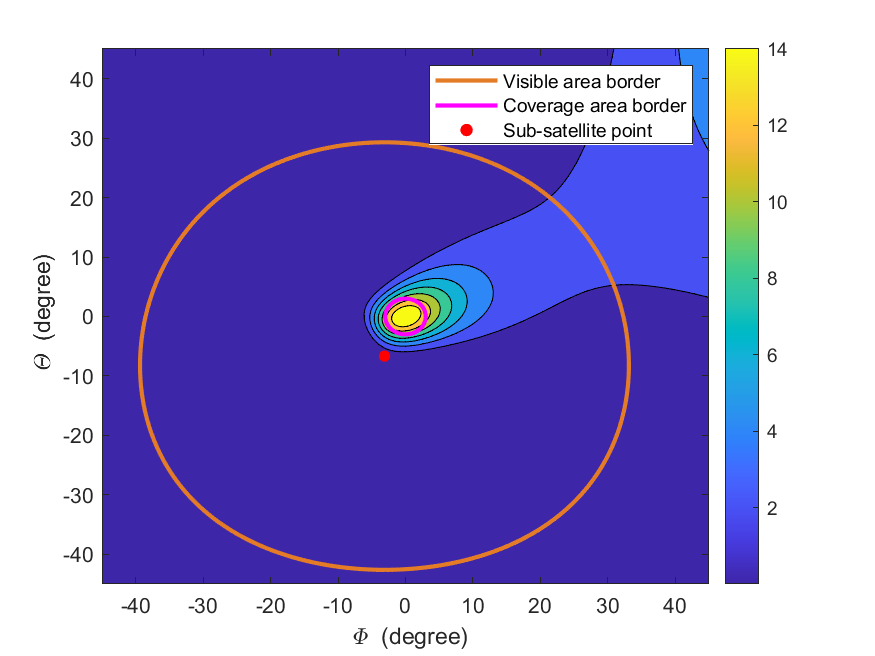} \label{Fig_patt_UPA_str1}}
	\subfigure[UPA, steering, $m=25$]{\includegraphics[width=5 cm]{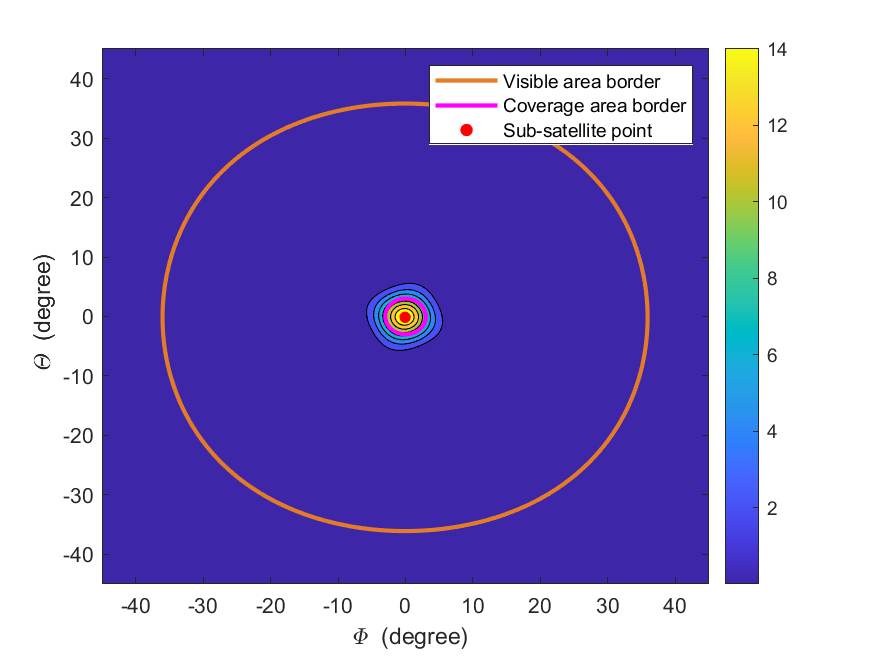} \label{Fig_patt_UPA_str25}}
	\subfigure[UPA, steering, $m=50$]{\includegraphics[width=5 cm]{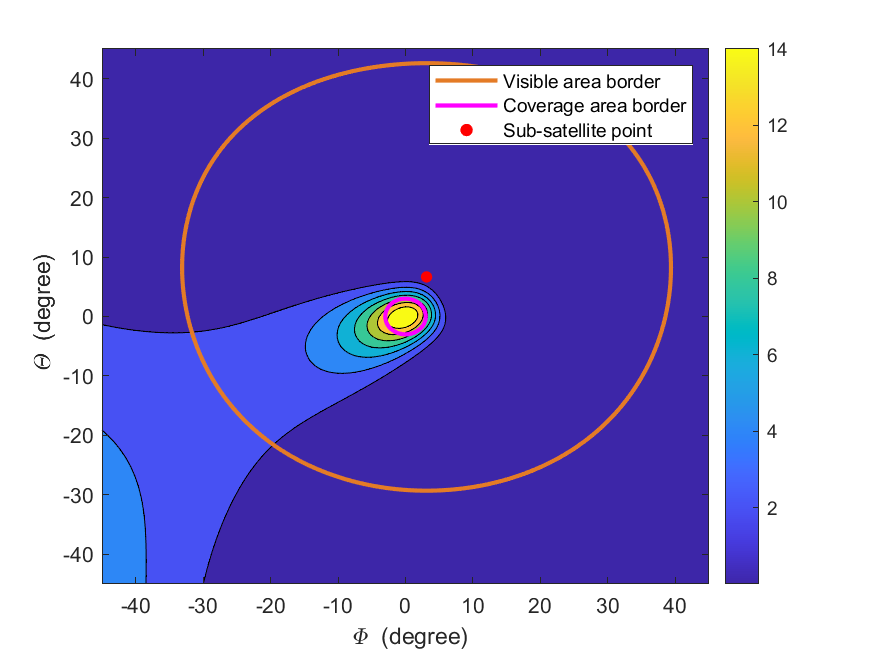} \label{Fig_patt_UPA_str50}}
	\subfigure[UPA, optimized, $m=1$]{\includegraphics[width=5 cm]{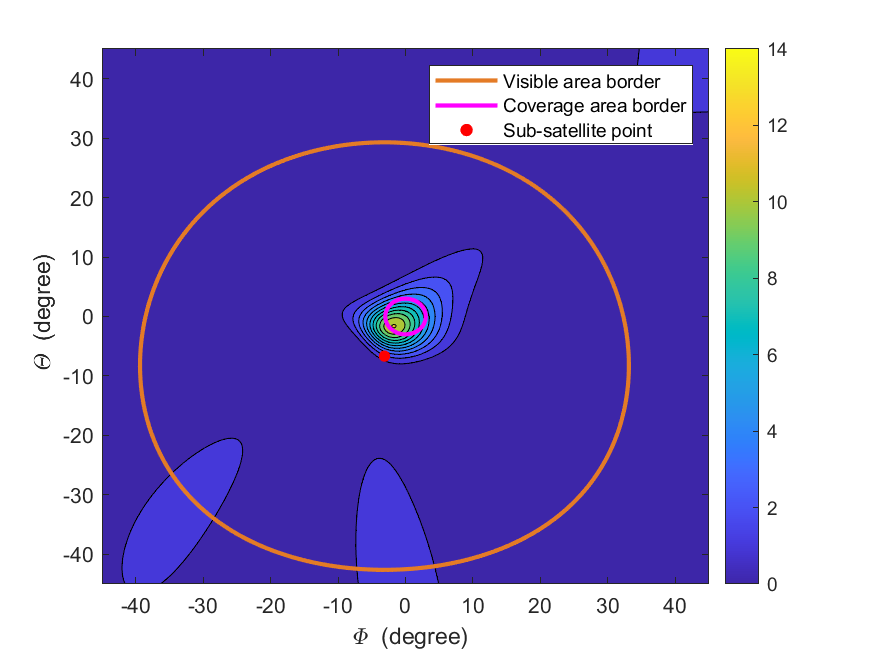} \label{Fig_patt_UPA_opt1}}
	\subfigure[UPA, optimized, $m=25$]{\includegraphics[width=5 cm]{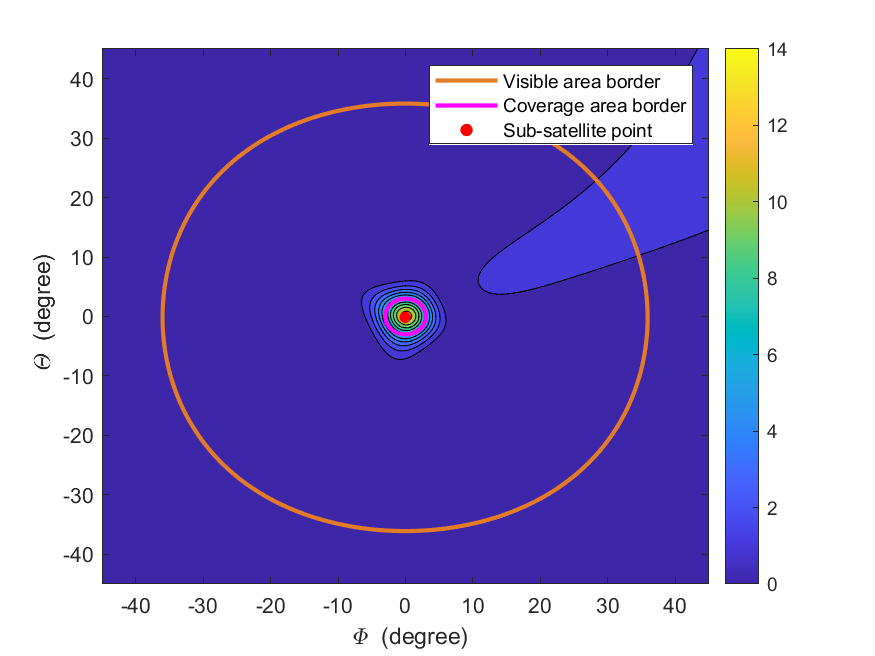} \label{Fig_patt_UPA_opt25}}
	\subfigure[UPA, optimized, $m=50$]{\includegraphics[width=5 cm]{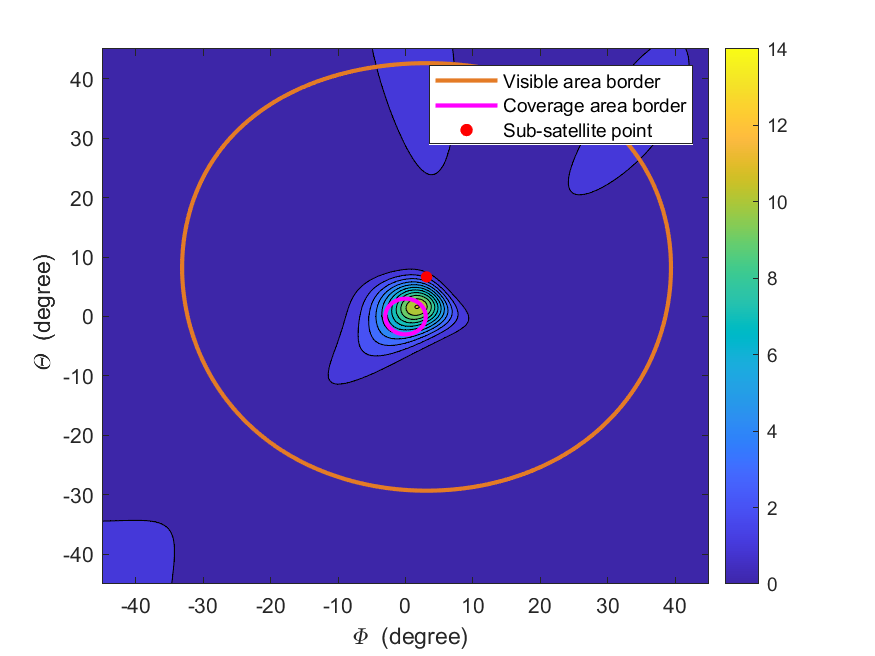} \label{Fig_patt_UPA_opt50}}
	\caption{Beamforming gain over the coverage and interference areas for the UPA-based schemes at different time slots.}
	\label{Fig_patt_UPA_all}
\end{figure*}

\begin{figure*}[t]
	\centering
	\subfigure[MA, $m=1$]{\includegraphics[width=5 cm]{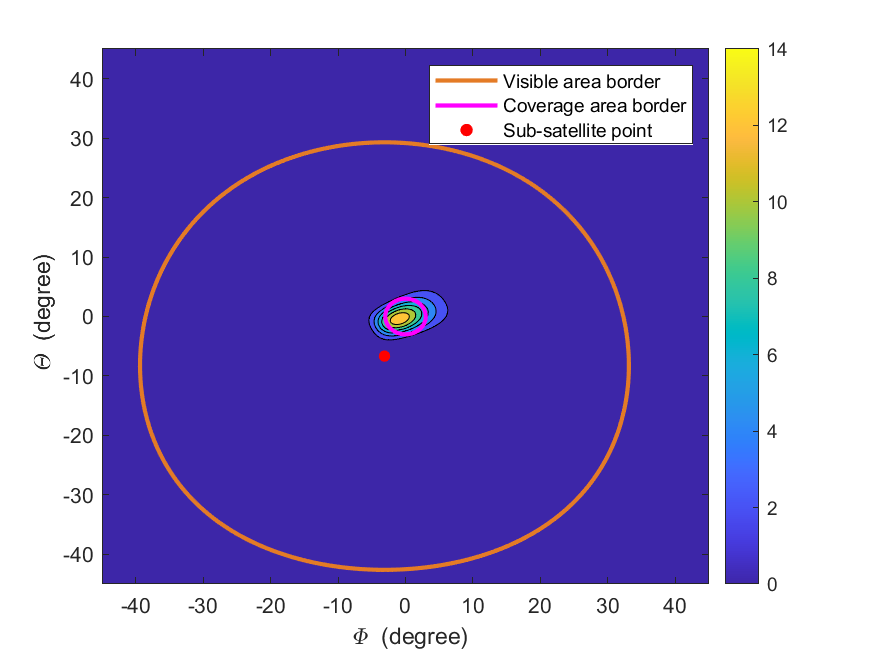} \label{Fig_patt_MA1}}
	\subfigure[MA, $m=25$]{\includegraphics[width=5 cm]{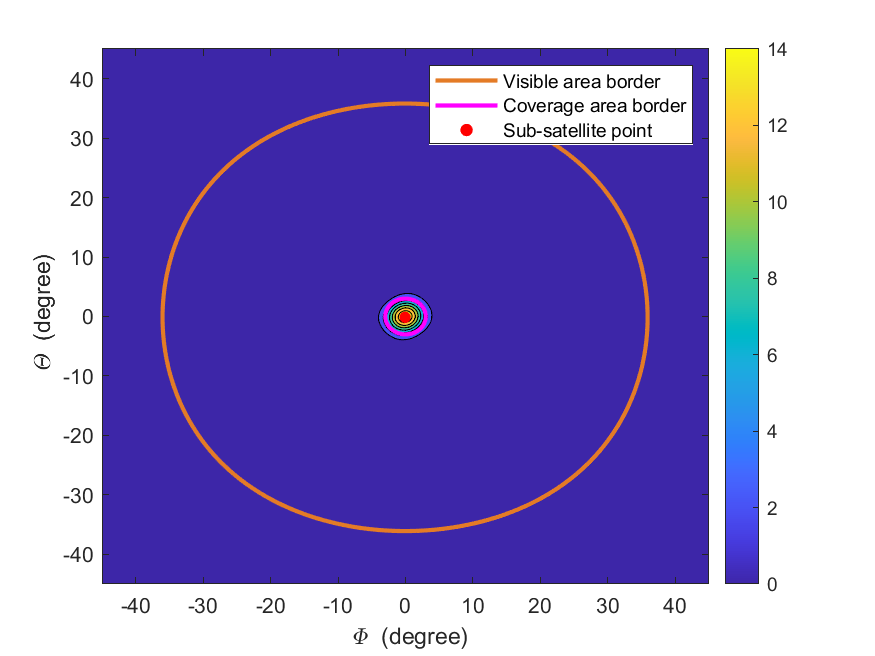} \label{Fig_patt_MA25}}
	\subfigure[MA, $m=50$]{\includegraphics[width=5 cm]{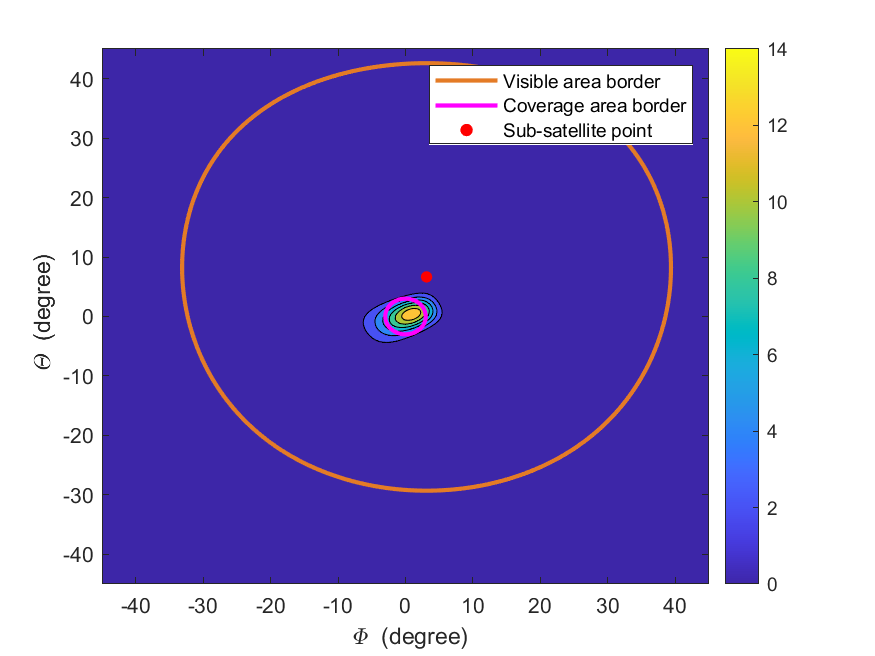} \label{Fig_patt_MA50}}
	\subfigure[LC-MA, $m=1$]{\includegraphics[width=5 cm]{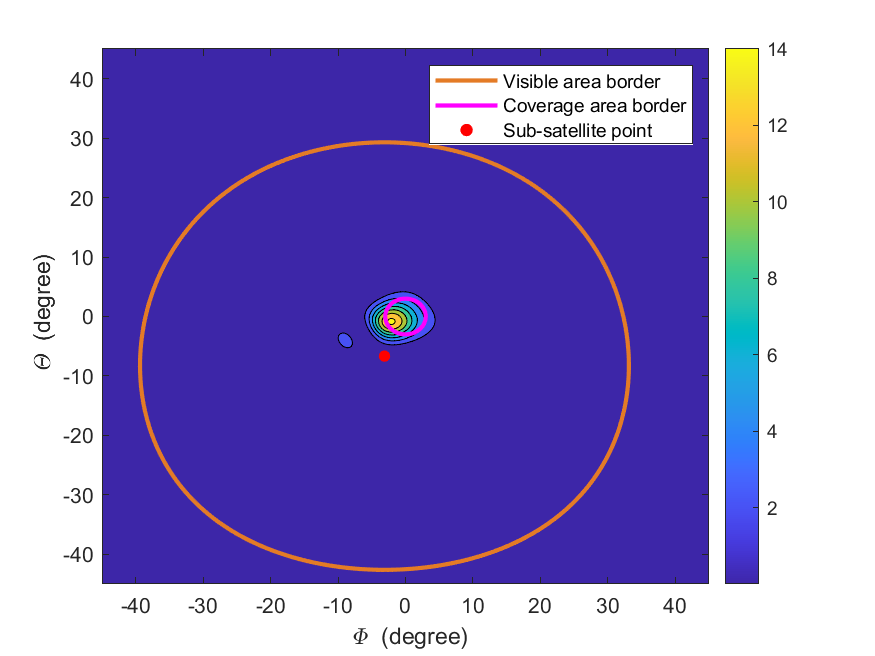} \label{Fig_patt_LC_MA1}}
	\subfigure[LC-MA, $m=25$]{\includegraphics[width=5 cm]{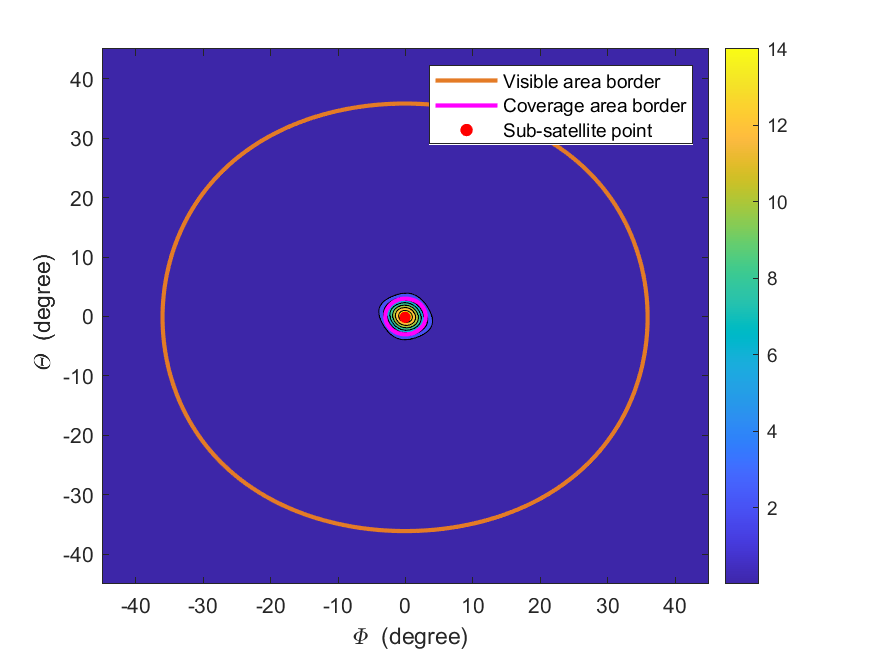} \label{Fig_patt_LC_MA25}}
	\subfigure[LC-MA, $m=50$]{\includegraphics[width=5 cm]{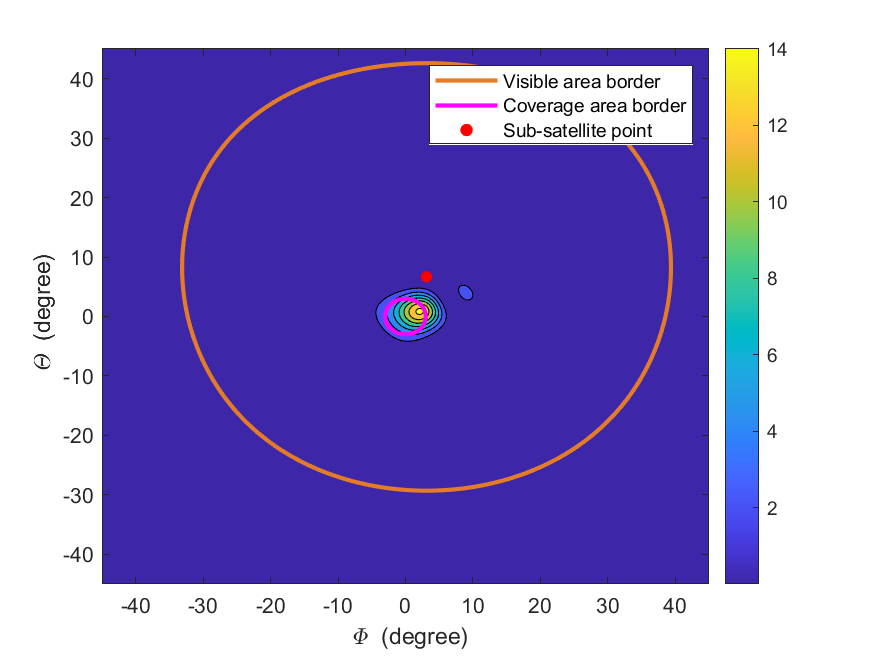} \label{Fig_patt_LC_MA50}}
	\caption{Beamforming gain over the coverage and interference areas for the MA and LC-MA schemes at different time slots.}
	\label{Fig_patt_MA_all}
\end{figure*}
First, we evaluate the convergence of the proposed AO-based algorithms and compare them with benchmark schemes in Fig. \ref{Fig_ite}. As observed, the proposed MA and LC-MA schemes both achieve a rapidly decreasing interference leakage power over the iterations subject to the same minimum beamforming gain over the coverage area, which validates the efficacy of the proposed AO-based algorithms. In particular, the MA and LC-MA algorithms converge after 800 and 200 iterations\footnote{The optimization can be implemented offline based on the given satellite orbit and coverage requirement.}, respectively, while the ratio of the converged interference power values between MA and LC-MA is 75\%. This result indicates that compared to the MA scheme, the LC-MA solution can significantly decrease the computational complexity and antenna movement overhead at the cost of partial interference mitigation gain. Moreover, we observe that the proposed MA schemes can yield significant performance gains over the UAP schemes in terms of interference leakage power reduction, which validates the advantage of MA arrays by dynamically reconfiguring the array geometry to adapt to the terrestrial coverage/interference area. In addition, the 6DMA scheme achieves a negligible performance gain over the MA scheme because the DoF in MAs' position optimization is much higher than that of the array orientation optimization for each time slot in the context of LEO satellite communications. However, the performance gap between the LC-MA and LC-6DMA schemes is observed to be larger. This is because although a common optimized array geometry is used for all time slots, the flexible array orientation in LC-6DMA can help further adjust the array responses over signal directions towards the interference area for reducing the interference leakage.

\begin{figure*}[t]
	\centering
	\subfigure[6DMA, $m=1$]{\includegraphics[width=5 cm]{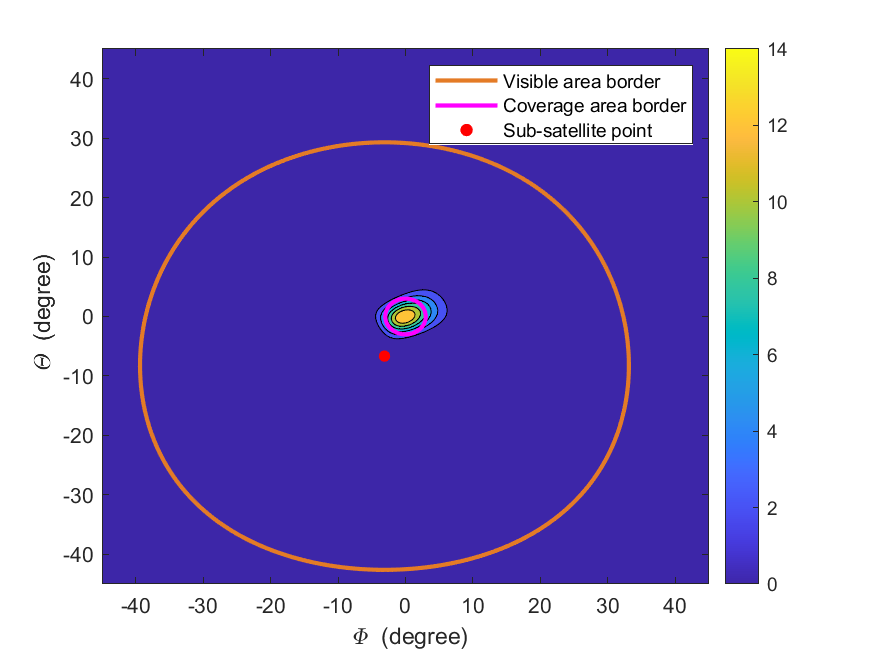} \label{Fig_patt_6DMA1}}
	\subfigure[6DMA, $m=25$]{\includegraphics[width=5 cm]{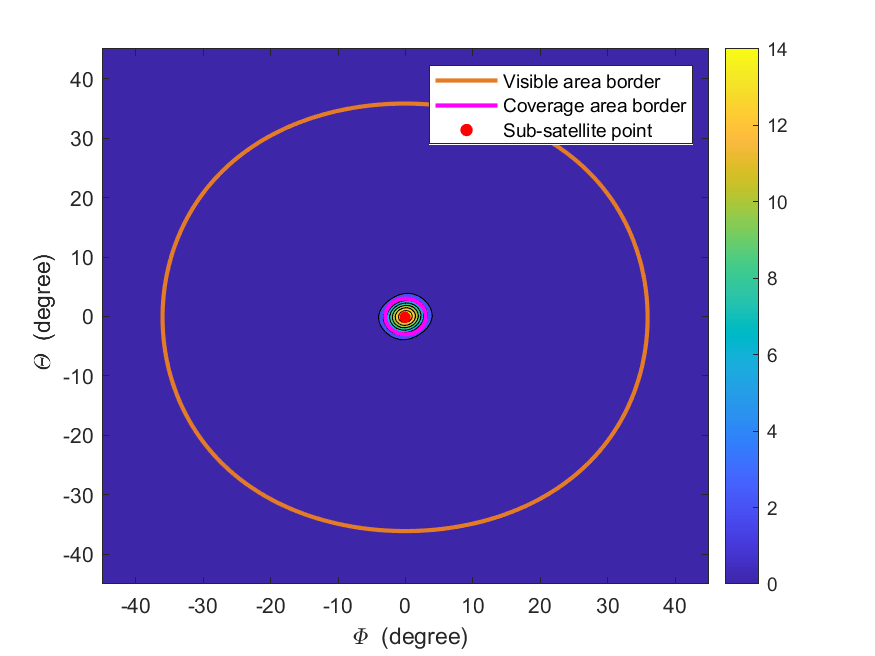} \label{Fig_patt_6DMA25}}
	\subfigure[6DMA, $m=50$]{\includegraphics[width=5 cm]{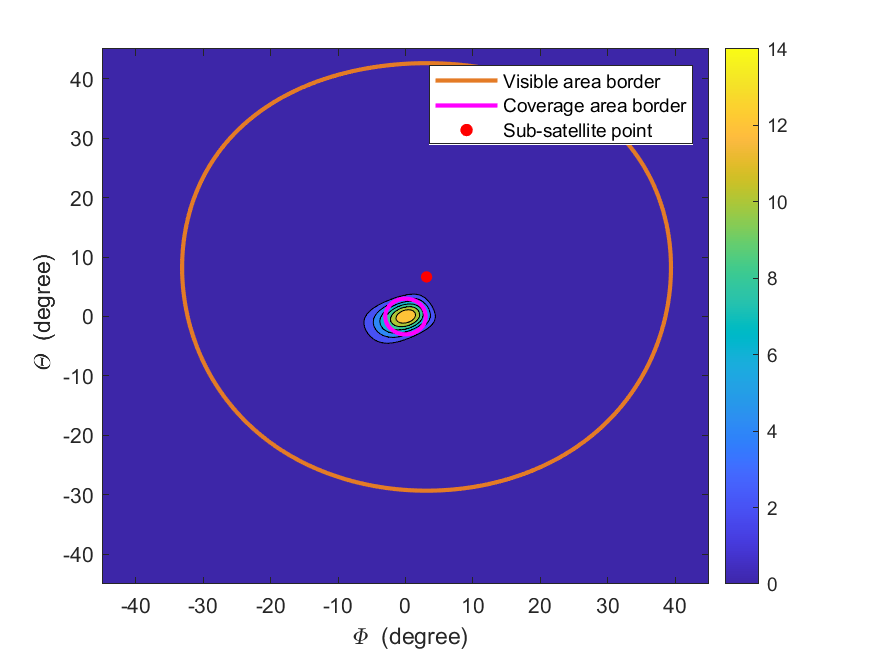} \label{Fig_patt_6DMA50}}
	\subfigure[LC-6DMA, $m=1$]{\includegraphics[width=5 cm]{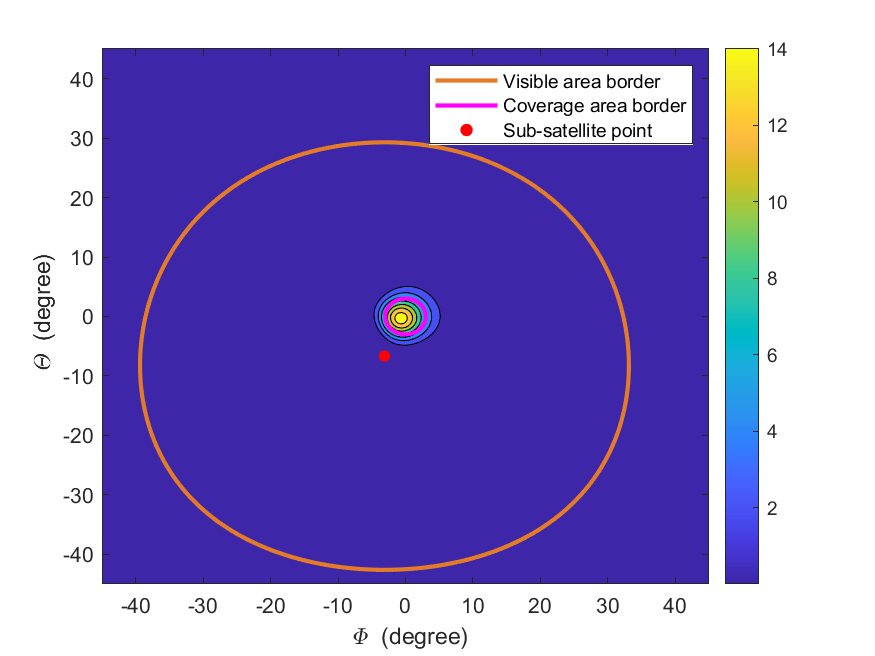} \label{Fig_patt_LC_6DMA1}}
	\subfigure[LC-6DMA, $m=25$]{\includegraphics[width=5 cm]{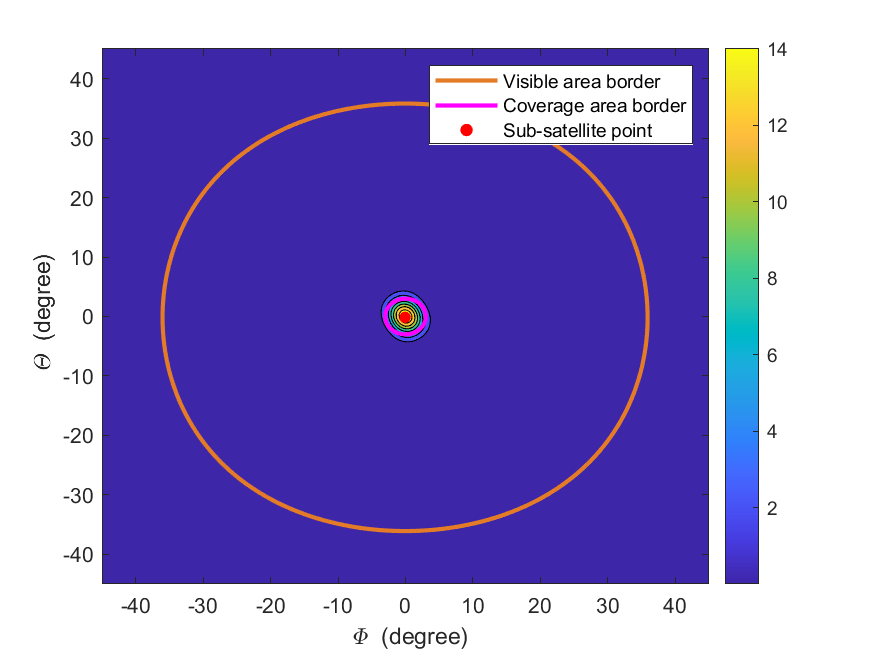} \label{Fig_patt_LC_6DMA25}}
	\subfigure[LC-6DMA, $m=50$]{\includegraphics[width=5 cm]{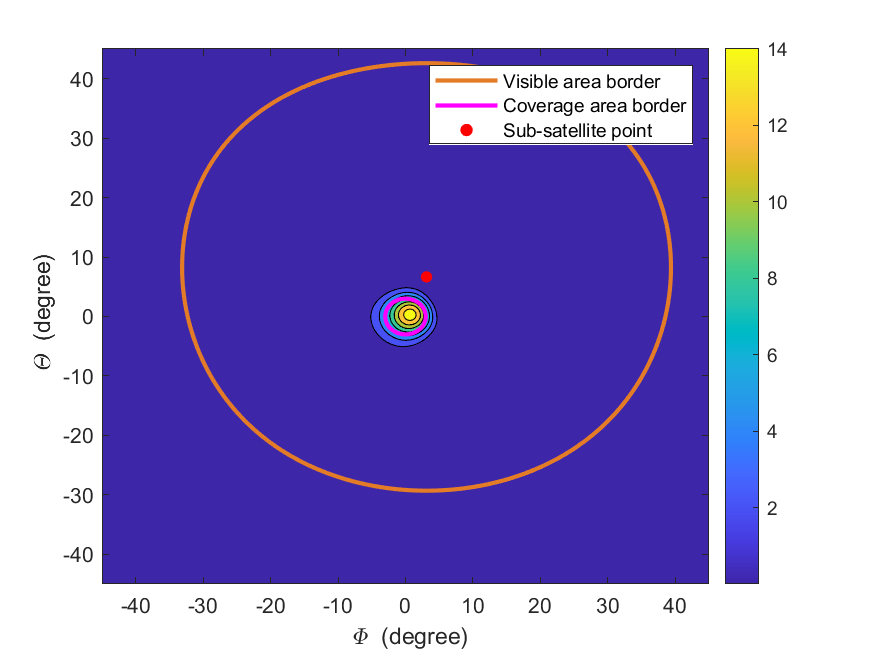} \label{Fig_patt_LC_6DMA50}}
	\caption{Beamforming gain over the coverage and interference areas for the 6DMA and LC-6DMA schemes at different time slots.}
	\label{Fig_patt_6DMA_all}
\end{figure*}

Next, we show in Fig. \ref{Fig_SLR} the average SLR achieved by the proposed and benchmark schemes, which is defined as the ratio of the average beamforming gain over all times slots to the average interference leakage power, i.e., $\frac{1}{M} \sum _{m=1}^{M} G_{m}(\mathbf{q}[m], \mathbf{w}[m])/I(\mathbf{q}, \mathbf{w})$. The proposed MA scheme can achieve about 5 dB gain in terms of SLR compared to the UPA-based schemes, which demonstrates the superiority of MA arrays in both coverage enhancement and interference mitigation. In addition, the SLR gap between the LC-MA/LC-6DMA scheme or MA/6DMA is about 1 dB, which shows the additional gain brought by array orientation/rotation. Moreover, it is observed that the MA and LC-6DMA schemes achieve a similar SLR performance. Since the array rotation is easier to implement compared to MA elements' movement, the LC-6DMA scheme provides a practically viable solution with the desired performance-complexity trade-off for the considered MA-aided satellite communication systems.

Figs. \ref{Fig_patt_UPA_all} and \ref{Fig_patt_MA_all} show the time-varying beam patterns of the UPA and MA schemes, respectively. Specifically, for each point $(\Theta, \Phi)$ on the earth surface, the beamforming gain $\left|\mathbf{a}(\mathbf{k}(\Theta, \Phi, t_m), \mathbf{q}(t_m))^{\mathrm{H}} \mathbf{w}(t_m)\right|^{2}$ in \eqref{eq_eff_channel_gain} is calculated at time slots $m=1,25,50$. The sub-satellite point refers to the position of the satellite projected on the earth surface. As can be observed, the beam patterns of UPA-based schemes have high leakage power to the interference area due to the strong sidelobe. In comparison, the proposed MA solution can flexibly configure beam patterns to decrease the sidelobe level by fully exploiting the DoFs in APV and AWV optimization. Besides, we can find that the beam patterns are symmetric at time slots $m=1$ and $m=50$ because the position of the satellite relative to the coverage area is also symmetric, as shown in the sub-satellite point in the figure. For $m=25$, the satellite locates right over the coverage center and thus the mainlobe has a circular shape. For $m=1$ and $m=50$, the satellite is located diagonally above the coverage area, which results in the distortion of the beam pattern w.r.t. the coverage area and thus the interference leakage increases. For the LC-MA strategy, the interference leakage becomes more severe because the array geometry cannot be reconfigured within the considered time interval. Nevertheless, it can still achieve significant performance improvement compared to the UPA-based schemes.

\begin{figure*}[t]
	\centering
	\subfigure[MA, $m=1$]{\includegraphics[width=4 cm]{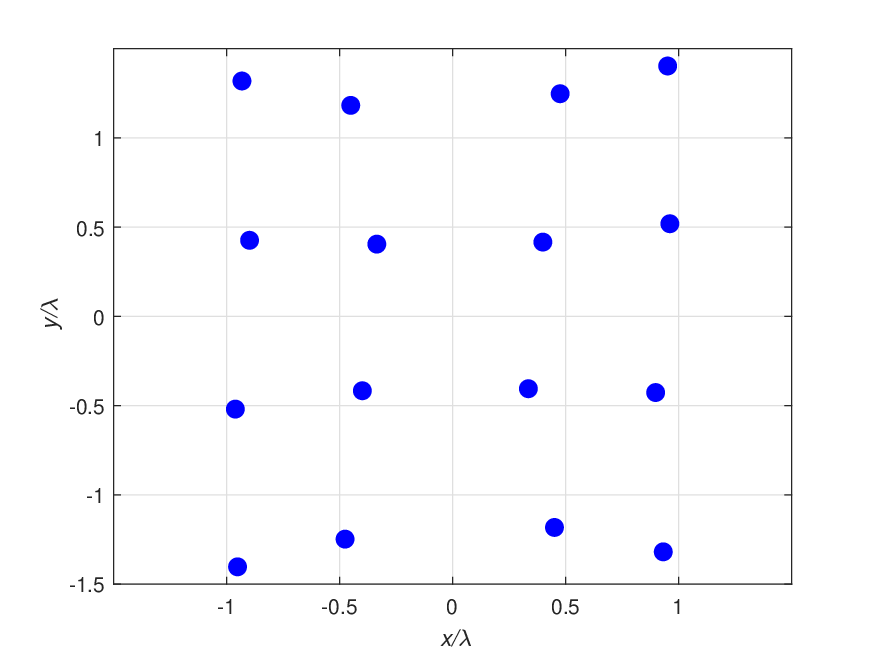} \label{Fig_geo_MA1}}
	\subfigure[MA, $m=25$]{\includegraphics[width=4 cm]{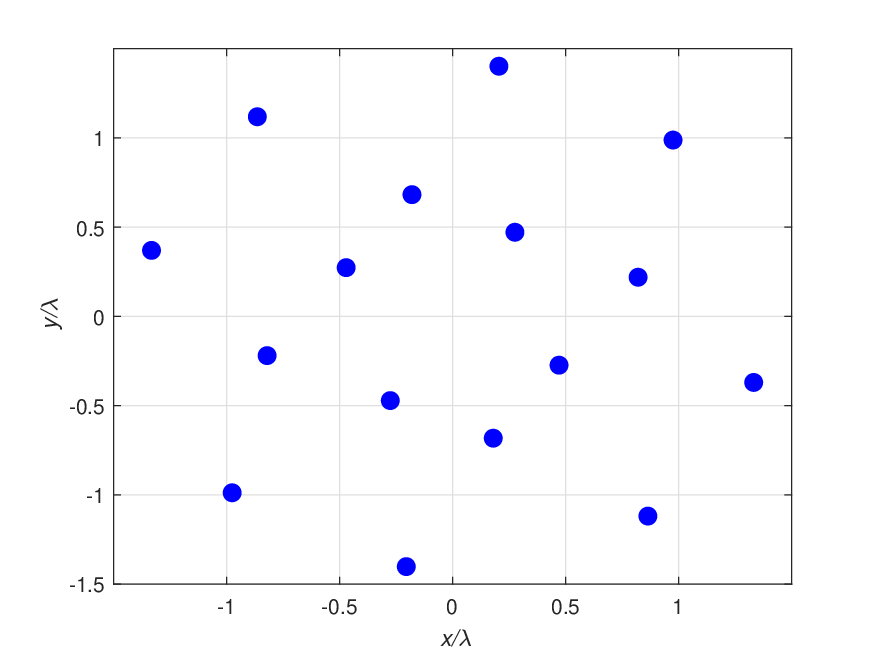} \label{Fig_geo_MA25}}
	\subfigure[MA, $m=50$]{\includegraphics[width=4 cm]{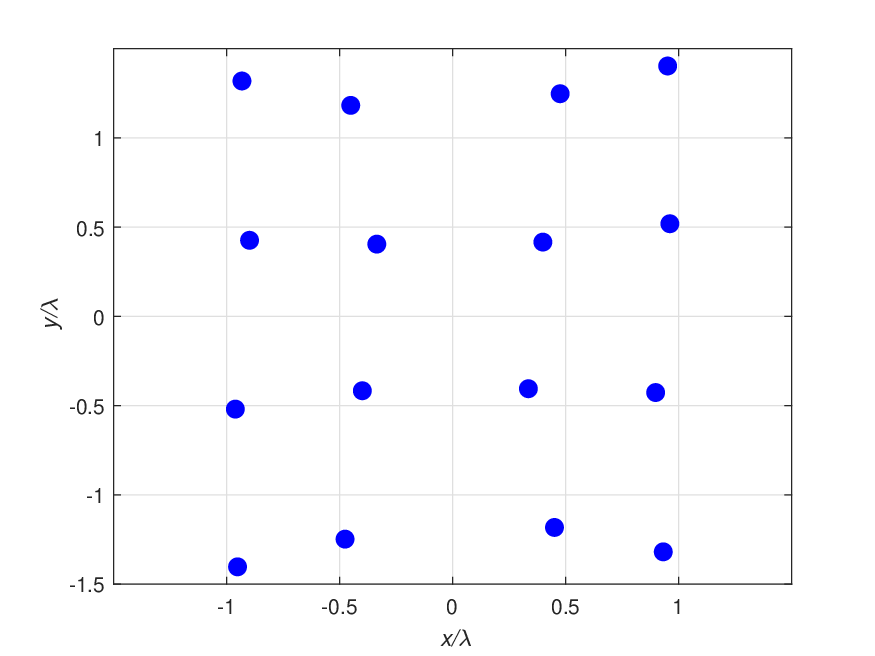} \label{Fig_geo_MA50}}
	\subfigure[LC-MA]{\includegraphics[width=4 cm]{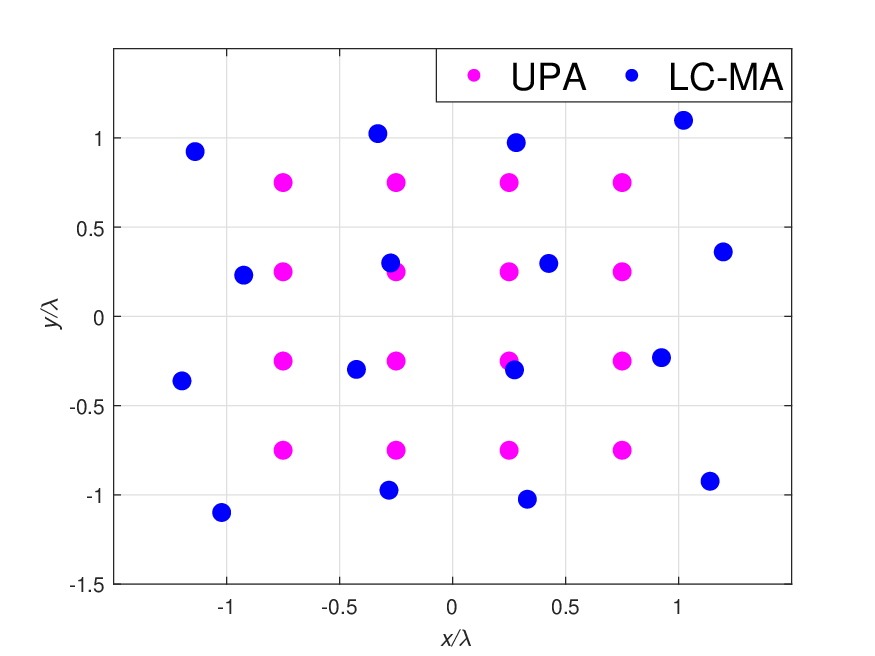} \label{Fig_geo_LC_MA}}
	\subfigure[6DMA, $m=1$]{\includegraphics[width=4 cm]{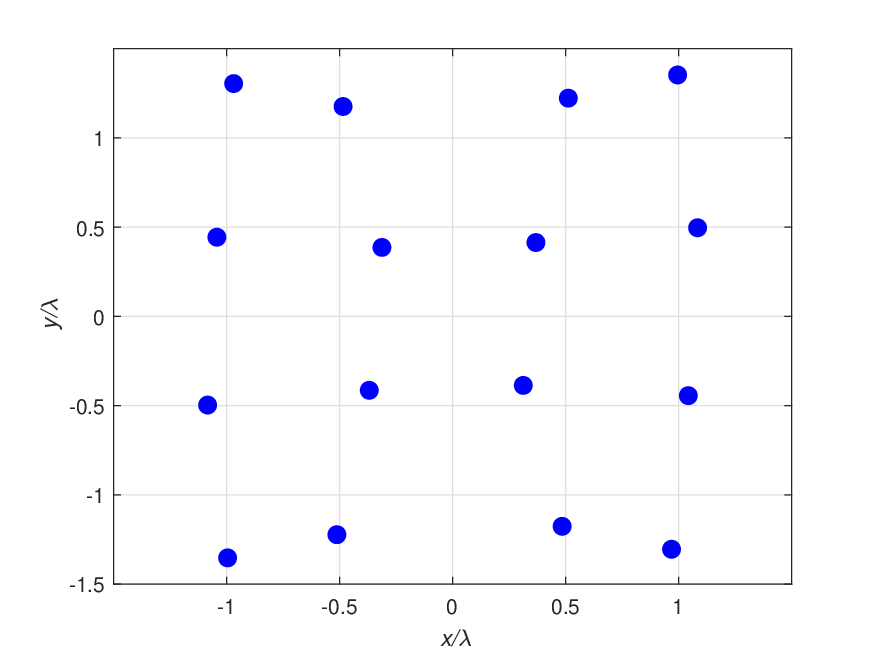} \label{Fig_geo_6DMA1}}
	\subfigure[6DMA, $m=25$]{\includegraphics[width=4 cm]{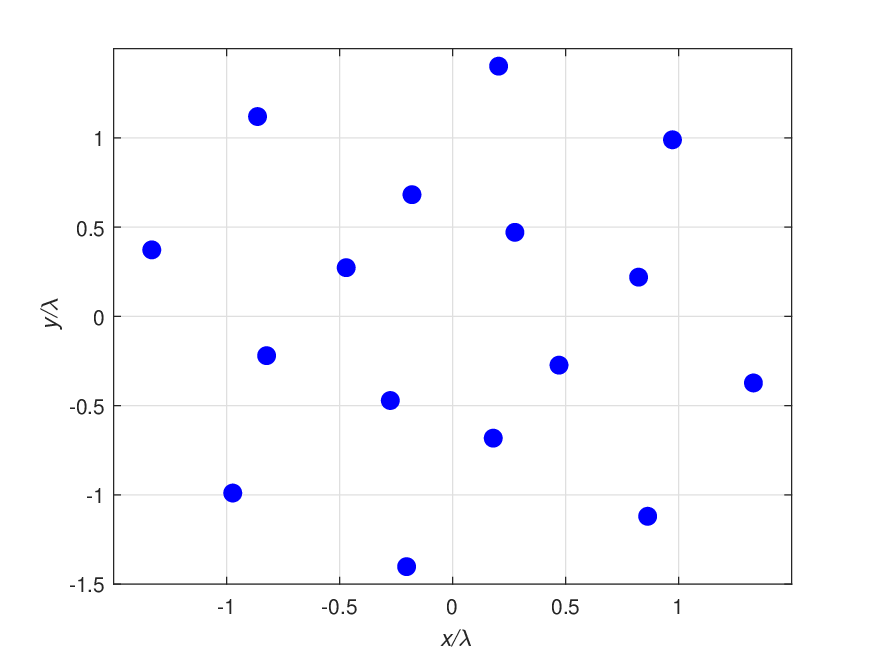} \label{Fig_geo_6DMA25}}
	\subfigure[6DMA, $m=50$]{\includegraphics[width=4 cm]{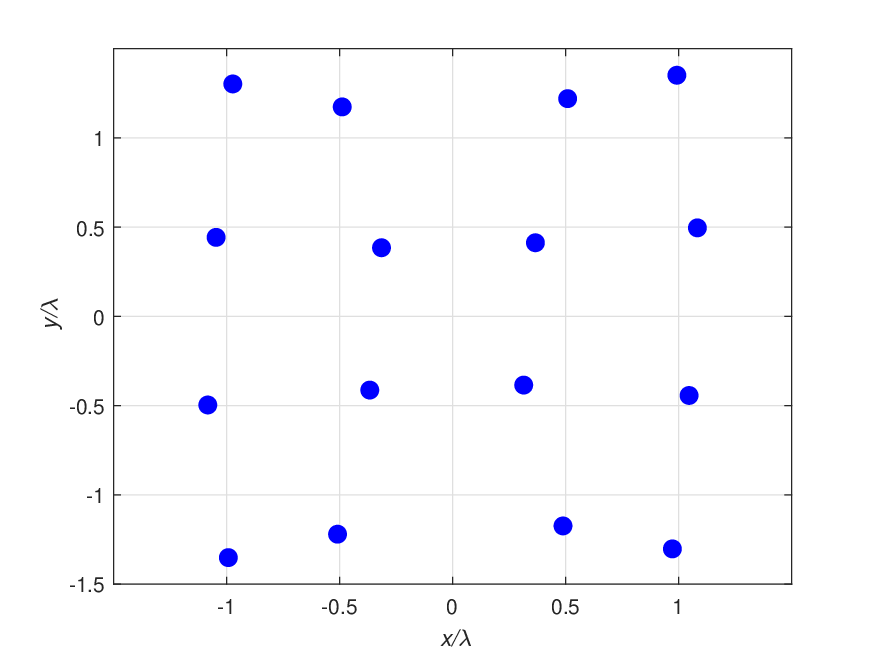} \label{Fig_geo_6DMA50}}
	\subfigure[LC-6DMA]{\includegraphics[width=4 cm]{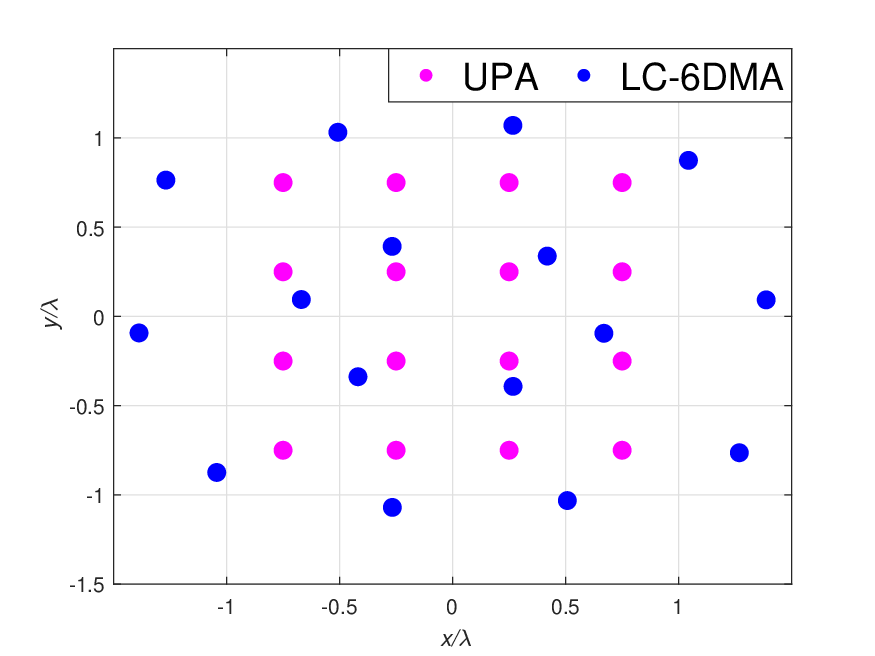} \label{Fig_geo_LC_6DMA}}
	\caption{The designed array geometry for the proposed MA, LC-MA, 6DMA, and LC-6DMA schemes at different time slots.}
	\label{Fig_geo_MA_all}
\end{figure*}

\begin{figure}[t]
	\centering
	\includegraphics[width=\figwidth cm]{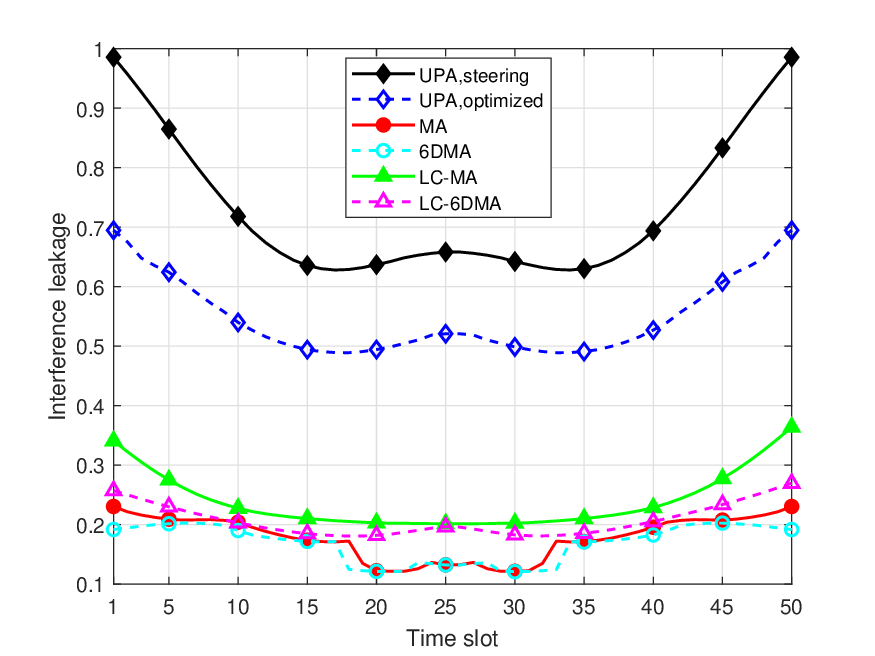}
	\caption{Comparison of the average interference leakage power versus the time slots.}
	\label{Fig_time}
\end{figure}

In Fig. \ref{Fig_patt_6DMA_all}, we show the time-varying beam patterns of the 6DMA-based schemes by considering additional optimization of the MA array orientation. For $m=25$, 6DMA and MA achieve similar beam patterns because the MA array orientates to the coverage center for both schemes. However, for time slots $m=1$ and $m=50$, the 6DMA scheme can alleviate the beam distortion compared to the MA scheme and thus the interference leakage power can be further reduced by optimizing the array orientation. Moreover, although the LC-6DMA scheme leads to a higher interference leakage power than the 6DMA scheme, its sidelobe level can be suppressed compared to the LC-MA scheme.

To shed more light on the antenna position optimization with or without orientation adjustment, we show in Fig. \ref{Fig_geo_MA_all} the array geometry (i.e., the optimized APV) for the MA, LC-MA, 6DMA, and LC-6DMA schemes. It can be observed that the MA and 6DMA schemes have similar array geometry for all time slots, and the array orientation optimization cannot attain much performance gain in decreasing the average interference leakage power. In contrast, the array geometries and orientations are different for LC-MA and LC-6DMA schemes, which result in distinguished interference mitigation performance. It is worth noting that compared to the UPA with half-wavelength inter-antenna spacing (shown in Figs. \ref{Fig_geo_LC_MA} and \ref{Fig_geo_LC_6DMA} for comparison), the optimized common array geometry for LC-MA/LC-6DMA has a larger array aperture and non-uniform antenna spacing, which can achieve a beam with a narrower main lobe and lower sidelobes for more effective interference mitigation, as shown in Figs. \ref{Fig_patt_UPA_all}, \ref{Fig_patt_MA_all}, and \ref{Fig_patt_6DMA_all}. For MA/6DMA at $m=25$, since the satellite is located right over the coverage center, the optimized MA array has an approximately circular geometry such that the beam pattern can exactly match the circular coverage area on the earth and reduce the beam leakage to the interference area. In addition, the satellite orbit and the coverage area are both centrosymmetric w.r.t. the coverage center, and thus the optimized array geometries at $m=1$ and $m=50$ for MA/6DMA exhibit rotational symmetry.

In Fig. \ref{Fig_time}, we show the interference leakage power within the considered time interval, which is quantized into $M=50$ time slots. It can be observed again that the proposed MA-based schemes always outperform UPA-based schemes during the considered time interval. In particular, for time slots 15-35, the satellite is nearly right above the coverage area and the interference leakage for all schemes is relatively small. However, for time slots 1-15 and 35-50, since the satellite is located diagonally above the coverage area, it is more challenging to mitigate the interference leakage due to the beam deformation. Nevertheless, the proposed MA and 6DMA schemes can still achieve low interference leakage at these time slots by flexibly reconfiguring the array geometry/orientation. In comparison, the LC-MA and LC-6DMA result in slightly higher interference leakage power at all times slots as compared to MA and 6DMA, respectively, due to the limited DoF in antenna movement optimization.

\vspace{-0.3 cm}
\section{Conclusion}
\begin{spacing}{0.9}
	In this paper, we investigated MA array enhanced beam coverage and interference mitigation for LEO satellite communication systems. Given the satellite orbit and terrestrial coverage requirement within a specific time interval, the APV and AWV of the satellite-mounted MA array were jointly optimized over time to minimize the average signal leakage power to the interference area of the satellite, subject to the constraints of the minimum beamforming gain over the coverage area, the feasible moving region of MAs, the minimum inter-MA spacing, the maximum moving speed of MAs, and the constant modulus of AWV. To address this non-convex problem involving continuous-time decision process, we first transformed it into a more tractable discrete-time optimization problem by discretizing the time interval and the angular coordinate of terrestrial areas. Then, an AO-based algorithm was developed by iteratively optimizing the APV and AWV at all discrete time slots, where the SCA technique was utilized to obtain locally optimal solutions over the iterations. To further reduce the antenna movement overhead, we proposed an LC-MA scheme by using an optimized common APV at all time slots. The positions of MAs can be reconfigured at the first time slot based on the coverage requirement, while subsequent antenna movement is dispensed until the coverage requirement of the satellite changes. Simulation results showed that compared to the conventional UPA-based schemes, the proposed MA array-aided beamforming schemes can significantly decrease the interference leakage and increase the average SLR of satellite-ground links. Moreover, the LC-MA scheme can achieve a performance comparable to the continuous-movement scheme in terms of interference mitigation, while additional orientation/rotation adjustment of the MA array (i.e., 6DMA) can further improve the performance, especially when the antenna position is not adjusted frequently.
\end{spacing}

\vspace{-0.3 cm}
\bibliographystyle{IEEEtran} 
\bibliography{IEEEabrv,ref_zhu}

\begin{thebibliography}{10}
\providecommand{\url}[1]{#1}
\csname url@samestyle\endcsname
\providecommand{\newblock}{\relax}
\providecommand{\bibinfo}[2]{#2}
\providecommand{\BIBentrySTDinterwordspacing}{\spaceskip=0pt\relax}
\providecommand{\BIBentryALTinterwordstretchfactor}{4}
\providecommand{\BIBentryALTinterwordspacing}{\spaceskip=\fontdimen2\font plus
\BIBentryALTinterwordstretchfactor\fontdimen3\font minus
  \fontdimen4\font\relax}
\providecommand{\BIBforeignlanguage}[2]{{%
\expandafter\ifx\csname l@#1\endcsname\relax
\typeout{** WARNING: IEEEtran.bst: No hyphenation pattern has been}%
\typeout{** loaded for the language `#1'. Using the pattern for}%
\typeout{** the default language instead.}%
\else
\language=\csname l@#1\endcsname
\fi
#2}}
\providecommand{\BIBdecl}{\relax}
\BIBdecl

\bibitem{dang2020should}
S.~Dang, O.~Amin, B.~Shihada, and M.-S. Alouini, ``What should {6G} be?''
  \emph{Nature Electronics}, vol.~3, no.~1, pp. 20--29, Jan. 2020.

\bibitem{wang2023road}
C.-X. Wang, X.~You, X.~Gao \emph{et~al.}, ``On the road to {6G}: Visions,
  requirements, key technologies, and testbeds,'' \emph{IEEE Commun. Surveys
  Tuts.}, vol.~25, no.~2, pp. 905--974, 2nd Quart. 2023.

\bibitem{ITU6G}
\BIBentryALTinterwordspacing
{ITU-R WP5D}, ``Future technology trends of terrestrial international mobile
  telecommunications systems towards 2030 and beyond,'' Nov. 2022. [Online].
  Available: \url{https://www.itu.int/pub/R-REP-M.2516}
\BIBentrySTDinterwordspacing

\bibitem{s21165284}
M.~Dangana, S.~Ansari, Q.~H. Abbasi, S.~Hussain, and M.~A. Imran, ``Suitability
  of {NB-IoT} for indoor industrial environment: A survey and insights,''
  \emph{Sensors}, vol.~21, no.~16, 2021.

\bibitem{Ghena2019challenge}
B.~Ghena, J.~Adkins, L.~Shangguan, K.~Jamieson, P.~Levis, and P.~Dutta,
  ``Challenge: Unlicensed {LPWANs} are not yet the path to ubiquitous
  connectivity,'' in \emph{International Conf. Mobile Comput. Netw.}, New York,
  NY, USA, Oct. 2019, pp. 1--12.

\bibitem{Abouzaid2021sky}
L.~Abouzaid, E.~Sabir, H.~Elbiaze, A.~Errami, and O.~Benhmammouch, ``The
  meshing of the sky: Delivering ubiquitous connectivity to ground internet of
  things,'' \emph{IEEE Internet Things J.}, vol.~8, no.~5, pp. 3743--3757, Mar.
  2021.

\bibitem{Di2019LEO}
B.~Di, L.~Song, Y.~Li, and H.~V. Poor, ``Ultra-dense {LEO}: Integration of
  satellite access networks into {5G} and beyond,'' \emph{IEEE Wireless
  Commun.}, vol.~26, no.~2, pp. 62--69, Apr. 2019.

\bibitem{Liu2021LEO}
S.~Liu, Z.~Gao, Y.~Wu, D.~W. Kwan~Ng, X.~Gao, K.-K. Wong, S.~Chatzinotas, and
  B.~Ottersten, ``{LEO} satellite constellations for {5G} and beyond: How will
  they reshape vertical domains?'' \emph{IEEE Commun. Mag.}, vol.~59, no.~7,
  pp. 30--36, Jul. 2021.

\bibitem{xiao2022LEO}
Z.~Xiao, J.~Yang, T.~Mao, C.~Xu, R.~Zhang, Z.~Han, and X.-G. Xia, ``{LEO}
  satellite access network ({LEO-SAN}) towards {6G}: Challenges and
  approaches,'' \emph{IEEE Wireless Commun.}, 2022, early access, DOI:
  10.1109/MWC.011.2200310.

\bibitem{Akan2023IoE}
O.~B. Akan, E.~Dinc, M.~Kuscu, O.~Cetinkaya, and B.~A. Bilgin, ``Internet of
  everything ({IoE}) - from molecules to the universe,'' \emph{IEEE Commun.
  Mag.}, vol.~61, no.~10, pp. 122--128, Oct. 2023.

\bibitem{Yahya2015antenna}
Y.~Rahmat-Samii and A.~C. Densmore, ``Technology trends and challenges of
  antennas for satellite communication systems,'' \emph{IEEE Tran. Antennas
  Propagat.}, vol.~63, no.~4, pp. 1191--1204, Apr. 2015.

\bibitem{Rao2015antenna}
S.~K. Rao, ``Advanced antenna technologies for satellite communications
  payloads,'' \emph{IEEE Tran. Antennas Propagat.}, vol.~63, no.~4, pp.
  1205--1217, Apr. 2015.

\bibitem{He2021review}
G.~He, X.~Gao, L.~Sun, and R.~Zhang, ``A review of multibeam phased array
  antennas as {LEO} satellite constellation ground station,'' \emph{IEEE
  Access}, vol.~9, pp. 147\,142--147\,154, 2021.

\bibitem{Moon2019phased}
S.-M. Moon, S.~Yun, I.-B. Yom, and H.~L. Lee, ``Phased array shaped-beam
  satellite antenna with boosted-beam control,'' \emph{IEEE Tran. Antennas
  Propagat.}, vol.~67, no.~12, pp. 7633--7636, Dec. 2019.

\bibitem{Ning2023widebeam}
B.~Ning, T.~Wang, C.~Huang, Y.~Zhang, and Z.~Chen, ``Wide-beam designs for
  terahertz massive {MIMO}: {SCA-ATP} and {S-SARV},'' \emph{IEEE Internet
  Things J.}, vol.~10, no.~12, pp. 10\,857--10\,869, June 2023.

\bibitem{Ran2023dual}
J.~Ran, Y.~Wu, C.~Jin, P.~Zhang, and W.~Wang, ``Dual-band multipolarized
  aperture-shared antenna array for {Ku-/Ka-}band satellite communication,''
  \emph{IEEE Tran. Antennas Propagat.}, vol.~71, no.~5, pp. 3882--3893, May
  2023.

\bibitem{Ning2023THzbeam}
B.~Ning, Z.~Tian, W.~Mei, Z.~Chen, C.~Han, S.~Li, J.~Yuan, and R.~Zhang,
  ``Beamforming technologies for ultra-massive {MIMO} in terahertz
  communications,'' \emph{IEEE Open J. Commun. Society}, vol.~4, pp. 614--658,
  Feb. 2023.

\bibitem{Vigan2010sparse}
M.~C. Viganö, G.~Toso, P.~Angeletti, I.~E. Lager, A.~Yarovoy, and
  D.~Caratelli, ``Sparse antenna array for earth-coverage satellite
  applications,'' in \emph{Proc. Fourth European Conf. Antennas Propagat.},
  Apr. 2010, pp. 1--4.

\bibitem{Bucci2014sparse}
O.~M. Bucci, T.~Isernia, S.~Perna, and D.~Pinchera, ``Isophoric sparse arrays
  ensuring global coverage in satellite communications,'' \emph{IEEE Tran.
  Antennas Propagat.}, vol.~62, no.~4, pp. 1607--1618, Apr. 2014.

\bibitem{Deng2020much}
R.~Deng, B.~Di, S.~Chen, S.~Sun, and L.~Song, ``Ultra-dense {LEO} satellite
  offloading for terrestrial networks: How much to pay the satellite
  operator?'' \emph{IEEE Trans. Wireless Commun.}, vol.~19, no.~10, pp.
  6240--6254, Oct. 2020.

\bibitem{Deng2021many}
R.~Deng, B.~Di, H.~Zhang, L.~Kuang, and L.~Song, ``Ultra-dense {LEO} satellite
  constellations: How many {LEO} satellites do we need?'' \emph{IEEE Trans.
  Wireless Commun.}, vol.~20, no.~8, pp. 4843--4857, Aug. 2021.

\bibitem{zhu2023MAMag}
L.~Zhu, W.~Ma, and R.~Zhang, ``Movable antennas for wireless communication:
  Opportunities and challenges,'' \emph{IEEE Commun. Mag.}, Oct. 16, 2023,
  early access, DOI: 10.1109/MCOM.001.2300212.

\bibitem{zhu2024historical}
L.~Zhu and K.-K. Wong, ``Historical review of fluid antenna and movable
  antenna,'' \emph{arXiv preprint arXiv:2401.02362}, 2024.

\bibitem{wong2022bruce}
K.-K. Wong, K.-F. Tong, Y.~Shen, Y.~Chen, and Y.~Zhang, ``Bruce lee-inspired
  fluid antenna system: Six research topics and the potentials for {6G},''
  \emph{Front. Comms. Net.}, vol.~3, no. 853416, pp. 1--31, Mar. 2022.

\bibitem{zhu2023MAarray}
L.~Zhu, W.~Ma, and R.~Zhang, ``Movable-antenna array enhanced beamforming:
  Achieving full array gain with null steering,'' \emph{IEEE Commun. Lett.},
  vol.~27, no.~12, pp. 3340--3344, Dec. 2023.

\bibitem{ma2024multi}
W.~Ma, L.~Zhu, and R.~Zhang, ``Multi-beam forming with movable-antenna array,''
  \emph{IEEE Commun. Lett.}, vol.~28, no.~3, pp. 697--701, Mar. 2024.

\bibitem{zhu2022MAmodel}
L.~Zhu, W.~Ma, and R.~Zhang, ``Modeling and performance analysis for movable
  antenna enabled wireless communications,'' \emph{IEEE Trans. Wireless
  Commun.}, Nov. 14, 2023, early access, DOI: 10.1109/TWC.2023.3330887.

\bibitem{wong2020limit}
K.~K. Wong, A.~Shojaeifard, K.-F. Tong, and Y.~Zhang, ``Performance limits of
  fluid antenna systems,'' \emph{IEEE Commun. Lett.}, vol.~24, no.~11, pp.
  2469--2472, Nov. 2020.

\bibitem{mei2024movable}
W.~Mei, X.~Wei, B.~Ning, Z.~Chen, and R.~Zhang, ``Movable-antenna position
  optimization: A graph-based approach,'' \emph{arXiv preprint
  arXiv:2403.16886}, 2024.

\bibitem{zhu2024wideband}
L.~Zhu, W.~Ma, Z.~Xiao, and R.~Zhang, ``Performance analysis and optimization
  for movable antenna aided wideband communications,'' \emph{arXiv preprint
  arXiv:2401.08974}, 2024.

\bibitem{ma2022MAmimo}
W.~Ma, L.~Zhu, and R.~Zhang, ``{MIMO} capacity characterization for movable
  antenna systems,'' \emph{IEEE Trans. Wireless Commun.}, vol.~23, no.~4, pp.
  3392--3407, Apr. 2024.

\bibitem{chen2023joint}
X.~Chen, B.~Feng, Y.~Wu, D.~W.~K. Ng, and R.~Schober, ``Joint beamforming and
  antenna movement design for moveable antenna systems based on statistical
  {CSI},'' in \emph{Proc. IEEE Global Commun. Conf.}, Kuala Lumpur, Malaysia,
  Dec. 2023, pp. 4387--4392.

\bibitem{zhu2023MAmultiuser}
L.~Zhu, W.~Ma, B.~Ning, and R.~Zhang, ``Movable-antenna enhanced multiuser
  communication via antenna position optimization,'' \emph{IEEE Trans. Wireless
  Commun.}, Dec. 12, 2023, early access, DOI: 10.1109/TWC.2023.3338626.

\bibitem{xiao2023multiuser}
Z.~Xiao, X.~Pi, L.~Zhu, X.-G. Xia, and R.~Zhang, ``Multiuser communications
  with movable-antenna base station: Joint antenna positioning, receive
  combining, and power control,'' \emph{arXiv preprint arXiv:2308.09512}, 2023.

\bibitem{wu2023movable}
Y.~Wu, D.~Xu, D.~W.~K. Ng, W.~Gerstacker, and R.~Schober, ``Movable
  antenna-enhanced multiuser communication: Optimal discrete antenna
  positioning and beamforming,'' in \emph{Proc. IEEE Global Commun. Conf.},
  Kuala Lumpur, Malaysia, Dec. 2023, pp. 7508--7513.

\bibitem{Wong2023opport}
K.-K. Wong, K.-F. Tong, Y.~Chen, Y.~Zhang, and C.-B. Chae, ``Opportunistic
  fluid antenna multiple access,'' \emph{IEEE Trans. Wireless Commun.},
  vol.~22, no.~11, pp. 7819--7833, Nov. 2023.

\bibitem{hu2024power}
G.~Hu, Q.~Wu, K.~Xu, J.~Ouyang, J.~Si, Y.~Cai, and N.~Al-Dhahir, ``Fluid
  antennas-enabled multiuser uplink: A low-complexity gradient descent for
  total transmit power minimization,'' \emph{IEEE Commun. Lett.}, 2024, early
  access, DOI: 10.1109/LCOMM.2024.3352664.

\bibitem{qin2024antenna}
H.~Qin, W.~Chen, Z.~Li, Q.~Wu, N.~Cheng, and F.~Chen, ``Antenna positioning and
  beamforming design for fluid antenna-assisted multi-user downlink
  communications,'' \emph{IEEE Wireless Commun. Lett.}, 2024, early access,
  DOI: 10.1109/LWC.2024.3360117.

\bibitem{cheng2023sum}
Z.~Cheng, N.~Li, J.~Zhu, X.~She, C.~Ouyang, and P.~Chen, ``Sum-rate
  maximization for movable antenna enabled multiuser communications,''
  \emph{arXiv preprint arXiv:2309.11135}, 2023.

\bibitem{Yang2024movable}
S.~Yang, W.~Lyu, B.~Ning, Z.~Zhang, and C.~Yuen, ``Flexible precoding for
  multi-user movable antenna communications,'' \emph{IEEE Wireless Commun.
  Lett.}, 2024, early access, DOI: 10.1109/LWC.2024.3372569.

\bibitem{wang2024multiuser}
C.~Wang, G.~Li, H.~Zhang, K.-K. Wong, Z.~Li, D.~W.~K. Ng, and C.-B. Chae,
  ``Fluid antenna system liberating multiuser {MIMO} for {ISAC} via deep
  reinforcement learning,'' \emph{IEEE Trans. Wireless Commun.}, 2024, early
  access, DOI: 10.1109/TWC.2024.3376800.

\bibitem{hu2024secure}
G.~Hu, Q.~Wu, K.~Xu, J.~Si, and N.~Al-Dhahir, ``Secure wireless communication
  via movable-antenna array,'' \emph{IEEE Signal Process. Lett.}, vol.~31, pp.
  516--520, Jan. 2024.

\bibitem{cheng2024secure}
Z.~Cheng, N.~Li, J.~Zhu, X.~She, C.~Ouyang, and P.~Chen, ``Enabling secure
  wireless communications via movable antennas,'' in \emph{Proc. IEEE
  International Conf. Acoust., Speech, Signal Processing}, Apr. 2024, pp.
  9186--9190.

\bibitem{tang2024secure}
J.~Tang, C.~Pan, Y.~Zhang, H.~Ren, and K.~Wang, ``Secure {MIMO} communication
  relying on movable antennas,'' \emph{arXiv preprint arXiv:2403.04269}, 2024.

\bibitem{ma2023MAestimation}
W.~Ma, L.~Zhu, and R.~Zhang, ``Compressed sensing based channel estimation for
  movable antenna communications,'' \emph{IEEE Commun. Lett.}, vol.~27, no.~10,
  pp. 2747--2751, Oct. 2023.

\bibitem{xiao2023channel}
Z.~Xiao, S.~Cao, L.~Zhu, Y.~Liu, B.~Ning, X.-G. Xia, and R.~Zhang, ``Channel
  estimation for movable antenna communication systems: A framework based on
  compressed sensing,'' \emph{IEEE Trans. Wireless Commun.}, 2024, early
  access, DOI: 10.1109/TWC.2024.3385110.

\bibitem{shao20246DMA}
X.~Shao, Q.~Jiang, and R.~Zhang, ``{6D} movable antenna based on user
  distribution: Modeling and optimization,'' \emph{arXiv preprint
  arXiv:2403.08123}, 2024.

\bibitem{shao2024discrete}
X.~Shao, R.~Zhang, Q.~Jiang, and R.~Schober, ``{6D} movable antenna enhanced
  wireless network via discrete position and rotation optimization,''
  \emph{arXiv preprint arXiv:2403.08123}, 2024.

\bibitem{montenbruck2002satellite}
O.~Montenbruck, E.~Gill, and F.~Lutze, \emph{Satellite orbits: models, methods,
  and applications}.\hskip 1em plus 0.5em minus 0.4em\relax New York City, USA:
  Springer, 2002.

\bibitem{boyd2004convex}
S.~Boyd and L.~Vandenberghe, \emph{Convex {O}ptimization}.\hskip 1em plus 0.5em
  minus 0.4em\relax Cambridge, U.K.: Cambridge Univ. Press, 2004.

\end{thebibliography}

\end{document}